\documentclass[letterpaper,aps,showpacs,floatfix,11pt,prc]{revtex4-1}
\usepackage{graphicx}
\usepackage{amsmath,amssymb,amsbsy,bm}
\usepackage{graphicx}
\usepackage{comment}
\usepackage{float}
\usepackage[colorlinks=true,linkcolor=blue,citecolor=blue,urlcolor=blue]{hyperref}

\begin{document}

\title{Evolution of Charge Fluctuations and Correlations in the Hydrodynamic Stage of Heavy Ion Collisions}
\author{Scott Pratt and Jane Kim}
\affiliation{Department of Physics and Astronomy and National Superconducting Cyclotron Laboratory\\
Michigan State University, East Lansing, MI 48824~~USA}
\author{Christopher Plumberg}
\affiliation{School of Physics and Astronomy, University of Minnesota,
Minneapolis, MN,~~55455,~USA}.
\date{\today}


\begin{abstract}
Charge fluctuations for a baryon-neutral quark-gluon plasma have been calculated in lattice gauge theory. These fluctuations provide a well-posed rigorous representation of the quark chemistry of the vacuum for temperatures above $T_c\gtrsim 155$ MeV. Due to the finite lifetime and spatial extent of the fireball created in relativistic heavy ion collisions, charge-charge correlations can only equilibrate for small volumes due to the finite time required to transport charge. This constraint leads to charge correlations at finite relative position that evolve with time. The source and evolution of such correlations is determined by the evolution of the charge fluctuation and the diffusion constant for light quarks. Here, calculations are presented for the  evolution of such correlations superimposed onto hydrodynamic simulations. Results are similar to preliminary measurements from STAR, but significant discrepancies remain.

\end{abstract}

\maketitle


\section{Introduction}
\label{sec:intro}

Perhaps the most defining characteristic of the quark-qluon plasma (QGP) is its wealth of charges, especially those carried by light quarks: up and down charge and strangeness. For temperatures greater than $T_c\sim$ 155 MeV, the gluons and up-down-strange quarks provide 52 degrees of freedom, 36 of which are carried by the $u$, $d$ and $s$ quarks, which come in three colors, two spins and with anti-particles. Given that such matter is net neutral, one cannot describe the chemistry by the average charge. Further, given that the quarks are strongly interacting, the net number of quarks (quarks plus antiquarks) is not well defined. For example, the contribution of quarks and anti-quarks to gluonic modes depends on the method in which it is probed. However, the fluctuation of the charge is both well-posed and reflects and describes the chemical composition of the plasma. Because it is defined in terms of conserved charges, it can be calculated unambiguously in lattice gauge theory. For the three light charges, $uds$, the charge fluctuation (also referred to as the charge susceptibility) in a neutral plasma is a three-by-three matrix,
\begin{eqnarray}
\chi_{ab}\equiv \frac{1}{V}\langle Q_a Q_b\rangle,
\end{eqnarray}
where $Q_a$ represents the up, down or strange charges, and $V$ is the volume. If the matter were to have a net charge, $\langle Q_a\rangle\ne 0$, the definition would be altered by replacing $Q_a$ with its fluctuation, $Q_a\rightarrow Q_a-\langle Q_a\rangle$. As stated above, the fluctuation measure provides insight into the chemical makeup of the plasma. If particles of species $h$, carrying charges $q_{ha}$, were good quasi-particles and uncorrelated with one another, the only correlations would be those between charges on the same particle. The susceptibility would then be
\begin{equation}
\chi_{ab}=\sum_h n_h q_{ha}q_{hb},
\end{equation}
where $n_h$ is the density of species $h$. For a gas of non-interacting u,d,s quarks, the susceptibility would be,
\begin{eqnarray}
\chi=\left(\begin{array}{ccc} n_u &0 &0\\
0& n_d &0\\
0 &0 & n_s\end{array}\right),
\end{eqnarray}
where $n_u, n_d$ and $n_s$ are the densities of up, down and strange quarks respectively. For a gas of hadrons a given species carries multiple charges, and $\chi$ develops off-diagonal elements. For example, pions or protons would provide non-zero $ud$ elements. If one had a gas of pions, where each pion species had density $n_\pi$, mixed with protons and anti-protons, both with densities $n_p$, the susceptibility would be
\begin{eqnarray}
\chi_{ab}=\left(\begin{array}{ccc}
8n_p & 2n_p-n_\pi & 0\\
2n_p-n_\pi & 2n_p & 0\\
0 & 0 &0\\
\end{array}\right).
\end{eqnarray}
The susceptibility from lattice calculations approaches that of a non-interacting gas of massless quarks at high temperature, and that of a hadron gas at temperatures below $T_c$. This is illustrated in Fig. \ref{fig:chilattice}, where the ratio $\chi/s$ is displayed as a function of the temperature $T$. The ratio is higher at low temperature because for a fixed amount of entropy both gases have similar numbers of particles, but the hadrons have multiple charges per particle. The transformation of behaving like a hadron gas to behaving like a quark-gluon plasma in the range $150 < T < 225$ MeV represents the fundamental change in degrees of freedom near and just above $T_c$. Given that the matter is strongly interacting, the fact that the susceptibilities closely resemble that of a non-interacting gas for $T>225$ MeV is remarkable, and suggests that inter-quark correlations are small in this temperature range, even if the collision rate is high.

In the most central collisions of heavy ions at the LHC or at RHIC, matter achieves temperatures greater than 300 MeV for times near 1.0 fm/$c$ after the initial collision \cite{Geiger:1992si,Geiger:1992ac,Dusling:2010rm,Gelis:2013rba}. The matter cools rapidly, staying in the QGP realm for the first $\sim 5$ fm/$c$, in the transition region for a few fm/$c$ then spends the last 10-25 fm/$c$ in the hadron phase before completely decoupling \cite{Bass:2000ib,Heinz:2013wva}. Unfortunately, measurements are confined to the outgoing particles, so inferring properties for a specific time or temperature is challenging. Even if measurements could be performed at a specific time, charge fluctuations would be limited in their ability to equilibrate due to the finite time for charges to spread. These limitations can be somewhat overcome by analyzing charge correlations, rather than charge fluctuations. Indexed by the spatial coordinate, charge correlations represent a more differential measure of the charge fluctuation,
\begin{eqnarray}
C_{ab}(\bm{r}_1,\bm{r}_2)&\equiv&\langle \rho_a(\bm{r}_1)\rho_b(\bm{r}_2)\rangle.
\end{eqnarray}
In subsequent sections, the measure of position, $\bm{r}$, will often be replaced by the spatial rapidity $\eta$.

In an equilibrated system, the correlation integrates to the charge fluctuation,
\begin{eqnarray}
\chi_{ab}(\bm{r}_1)&=&\int dr_2~C_{ab}(\bm{r}_1,\bm{r}_1+\bm{r}_2).
\end{eqnarray}
In a heavy-ion collision, the net charges are fixed and do not fluctuate. Thus, if the ``volume'' encompasses the entire collision the charge fluctuation is zero, and $C_{ab}$ integrates to zero. However, charge can diffuse away from small sub-volumes, and if the equilibrated correlations are confined to very short distances, the short-range correlation might readily equilibrate, leaving a residual longer-range correlation. When combined, the short-range correlation and the long-range correlation should integrate to zero due to charge conservation. We express the short-range correlation as a delta function,
\begin{eqnarray}
C_{ab}(\bm{r}_1,\bm{r}_2)&=&\chi_{ab}\delta(\bm{r}_2-\bm{r}_1)+C'_{ab}(\bm{r}_1,\bm{r}_2).
\end{eqnarray}
Here, $C'$ describes the correlation that would diffuse over large distances if given the chance, but even given large times would integrate to $-\chi$. If the short range correlations do not have time to equilibrate, $\chi$ could be replaced with some time-dependent function that would evolve toward $\chi$ if given sufficient time. For example, if the production of strangeness kept up with equilibrium in the plasma stage, one would expect $\chi_{ss}=n_s+n_{\bar{s}}$ to follow the equilibrated value. However, if chemical rates were slow, one might evolve $\chi_{ss}$ according to chemical rates rather than assuming the equilibrated value. The residual correlation is referred to as the ``balancing'' correlation, and ultimately leads to the measurement of the charge balance functions described here \cite{Bass:2000az,Pratt:2011bc}. At the end of the reaction, the short-range correlation, described by $\chi$, involves only those charges on the same hadron, and the strength of $C'(\Delta\bm{r},t)$ is constrained by charge conservation. Thus, it is only the spatial spread of $C'(\Delta\bm{r},t)$ that provides information. Because charge conservation is local, the spread $C'(\Delta\bm{r})$ indicates the times at where the source function fed the correlation, and the rate at which charge can diffuse or spread. The evolution of $C'$ is determined by charge conservation \cite{Pratt:2017lce},
\begin{eqnarray}
\label{eq:source}
D_t C'_{ab}(\bm{r}_1,\bm{r}_2,t)&=&-\bm{\nabla}_1\cdot\langle \bm{j}_a(\bm{r}_1,t)\rho_b(\bm{r}_2,t)\rangle
-\bm{\nabla}_2\cdot\langle \rho_a(\bm{r}_1,t)\bm{j}_b(\bm{r}_2,t)\rangle
-S_{ab}(\bm{r}_1,t)\delta(\bm{r}_2-\bm{r}_1),\\
\nonumber
S_{ab}(\bm{r},t)&=&D_t\chi_{ab}(\bm{r},t)+(\bm{\nabla}\cdot \bm{v})\chi_{ab}(\bm{r},t),
\end{eqnarray}
where $D_t$ is the co-moving derivative. If $\chi_{ab}$ is known as a function of the temperature, one can use the hydrodynamic evolution to extract the source function $S_{ab}(\bm{r},t)$. Then if one assumes the currents are determined diffusively, $\bm{j}_a=-D\bm{\nabla}\rho_a$, the evolution of the correlation function is determined.

For a quark gas, a positive source function corresponds to the creation of quarks, and a negative source function is caused by the annihilation of quarks. For an isentropic expansion of a massless parton gas, the number of quarks stays fixed, and $S_{ab}=0$. For a hadron gas, there are off-diagonal elements. For example, pions contribute negatively to $\chi_{ud}$ because they are constructed from $u\bar{d}$ or $d\bar{u}$ quarks. Thus, the off-diagonal elements of $S_{ab}$ tend to be negative. With the sign convention used in Eq. (\ref{eq:source}) the function $C'_{ab}$ is fed proportional to $-S_{ab}$ whereas the change is $\chi_{ab}$ is proportional to $+\chi_{ab}$.

Previous studies have also considered the growth of charge-balance correlations throughout time. Using a microscopic simulation based on hadronic degrees of freedom \cite{Cheng:2004zy}, it was seen that the width of such correlations in relative rapidity was found to be qualitatively inconsistent with data. In a purely hadronic scenario, there is little charge production after the initial production, and without a delayed late-time surge of the source function, models were found to be qualitatively inconsistent with data. In \cite{Pratt:2015jsa,Pratt:2011bc,Pratt:2012dz} the source function was parameterized as coming from two components, an early surge from the creation of charge in the equilibration of the QGP, and a second surge related to hadronization. Varying the strength of the surges and the diffusive spreads, and overlaying onto a blast-wave parametrization of the collective flow, experimental measurements were reasonably well matched when the strength of the initial surge of charge production was consistent with early chemical equilibration of the up, down and strange quarks. A similar parametric model, but with a more continuous description of the source function has also been attempted \cite{Pan:2015pzh}. Formalisms that can applied to three-dimensional hydrodynamic simulations have also been presented \cite{Ling:2013ksb,Pratt:2016lol}, but this study represents the first attempt at implementing such a formalism with a realistic hydrodynamic evolution.

An alternative method for evolving correlations in the context of a hydrodynamic evolution has been presented in \cite{Ling:2013ksb}. In these approaches, correlations are seeded by  hydrodynamical fluctuations, i.e. by adding stochastic source terms to the usual hydrodynamic equations of motion \cite{Kapusta:2011gt}.  These equations of motion are thereby converted into stochastic differential equations which must be solved by explicitly specifying correlation functions for the fluctuating source terms whose form is dictated by the fluctuation-dissipation theorem \cite{Kubo:1966}.  When these correlation functions are chosen proportional to $\delta$-functions in coordinate separation, the hydrodynamical fluctuations are termed ``white noise."  Although simple to implement in dynamically evolving systems such as heavy-ion collisions, systems with white noise can be shown to exhibit violations of relativistic causality, with disturbances propagating faster than the speed of light \cite{Kapusta:2014dja, Kapusta:2017hfi}.  One way of avoiding this problem is by replacing white noise with ``colored noise," i.e., by replacing coordinate-space $\delta$-functions with other functional forms which are characterized by finite widths in space and/or time \cite{Kapusta:2017hfi}. A second challenge for stochastic treatments concerns projecting the correlations onto hadrons. The stochastic correlations include the correlation of a particle with itself, and because these correlations are also reproduced in the hadronization of an uncorrelated background, one must be careful to remove that part of the correlation from the stochastic treatment.

In this study, we take a different approach to tackling these same problems of relativistically causal diffusion and the subtraction of hadronic self-correlations.  Instead of modifying the hydrodynamic equations of motion themselves, we retain them as non-stochastic (``smooth'') partial differential equations which can be solved by standard hydrodynamics codes \cite{Shen:2014vra}, and treat the stochastic component of the correlation by directly producing pairs of oppositely charged particles on top of the smooth, hydrodynamical evolution of the system, according to Eq. \eqref{eq:source}. These pairs provide a Monte-Carlo representation of the correlation function. Rather than having these pairs drift apart according to the diffusion equation, they instead move according to a random walk, moving at the speed of light, and punctuated by collisions which then randomize the direction of the velocity in the local matter frame. By choosing the collision rate appropriately, this evolution is consistent with the diffusion equation after many collisions, but differs when the number of collisions is small, or for times shortly after the pair is created. For these times, the diffusive separation is effectively cut off at large separation by the constraint of charges not moving faster than the speed of light. With the approach taken here, which is also discussed in \cite{Pratt:2016lol}, the discreteness of the charges can be treated in such a way that the correlation of hadrons with themselves, or those charges on a specific hadron with themselves, are not double-counted.  In the noise-based algorithms described above, those correlations are also included in the correlation function, which makes it difficult to build a hadronization sampling routine that does not redundantly produce such correlations. By requiring that individual particles travel no faster than the speed of light, we can enforce the requirement of relativistic causality without needing to modify the usual way in which the hydrodynamic equations are solved.  Instead, the diffusive dynamics are obtained by directly solving Eq. \eqref{eq:source}, and evolving pairs of charges as prescribed by $C'_{ab}$.

Stated more precisely, our principal goal in this study is to implement the following procedure for studying charge balance functions in heavy-ion collisions.  Beginning with the hydrodynamic solution for central $Au+Au$ collisions at energies corresponding to $\sqrt{s}_{nn}=200$ GeV, the highest energy for which data from RHIC (Relativistic Heavy-Ion Collider at Brookhaven National Laboratory) is available, source functions are extracted from the space-time history of the hydrodynamic evolution. The source functions feed the correlations functions, $C'_{ab}$, which are then evolved as functions of relative spatial rapidity according to the diffusion constant $D$. Both the charge susceptibility, $\chi_{ab}$ \cite{Borsanyi:2011sw,Bellwied:2015lba}, and the diffusion constant, $D$ \cite{Aarts:2014nba}, are taken from lattice calculations. The correlations are evolved until the hydrodynamic fluid elements reach a breakup temperature $T_f$, which is near or below $T_c$. At this point, the correlations are projected onto hadrons, using the assumption that small up, down and strange charges are distributed amongst the various hadron species thermodynamically. A method for describing the additional number of hadrons of a given species, $\delta N_h$, due to a small charge, $\delta Q_a$ was found in \cite{Pratt:2011bc},
\begin{eqnarray}
\delta N_h&=&\langle n_h\rangle q_{ha}\chi^{-1}_{ab}\delta Q_b,
\end{eqnarray}
where $\langle n_h\rangle$ is the equilibrium density of such hadrons at the hyper-surface, and $q_{ha}$ is the charge of type $a$ on a hadron of species $h$. Assuming thermal emission from the Cooper-Frye surface, the additional hadrons, $\delta N_h$, are then generated. After incorporating their thermal motion, plus decays, the correlations in coordinate space, $C'_{ab}(\bm{r}_1,\bm{r}_2)$ which are indexed by their $u,d$ and $s$ charges, are mapped onto correlations in momentum space of hadrons, $C'_{hh'}(\bm{p}_1,\bm{p}_2)$, indexed by the hadronic species. 

The hadronic correlations in momentum space are represented by ``generalized charge balance functions'', $B_{hh'}$, and are binned as functions of the relative rapidity in this study. Balance functions are defined by any two hadron species, $h$ and $h'$,
\begin{eqnarray}
\label{eq:baldef}
B_{hh'}(\Delta y)&=&\frac{1}{2\langle N_h\rangle}\int dydy'~\langle N_h(y)[N_{\bar{h}'}(y')-N_{h'}(y')]\rangle
\delta(\Delta y-|y-y'|)\\
\nonumber
&+&\frac{1}{2\langle N_{\bar{h}}\rangle}\int dydy'~\langle N_{\bar{h}}(y)[N_{h'}(y')-N_{\bar{h}'}(y')]\rangle
\delta(\Delta y-|y-y'|).
\end{eqnarray}
Here, $\langle N_h\rangle$ is the average number of species of type $h$, and $\langle N_h(y)N_{h'}(y')\rangle$ is the differential probability of observing species of type $h$ and $h'$ with rapdidities $y$ and $y'$ in the same event. The first term represents the conditional probability for observing hadrons of type $\bar{h}'$ vs $h'$ given the observation of a hadron $h$, with relative rapidity $\Delta y$. The numerators are constructed experimentally by summing over all pairs of particles where one is of type $h$ and the other is of type $h'$ and creating a histogram binned by relative rapidity. The histogram is incremented by unity for the combinations $h\bar{h}'$ and $\bar{h}h'$ and by $-1$ for $hh'$ or $\bar{h}\bar{h}'$. The sums ignore the terms where the pair refers to a particle with itself. These correlations have been measured by STAR for $pp$, $pK$, $KK$ and $\pi\pi$ \cite{Wang:2012jua}. They have also been measured by STAR, NA49 and by ALICE for the case where $h$ and $h'$ refer to hadron indexed by charge only \cite{Adamczyk:2015yga,Li:2011zzx,Abelev:2010ab,Alt:2007hk,Adams:2003kg,Aggarwal:2010ya,Abelev:2013csa,Alt:2004gx}. Here, we present calculations for all combinations of charged pions, charged kaons and protons. Results are projected through an acceptance filter provided by STAR and compared to their results. 

The final step is to identify those free parameters within our model which may be tuned to obtain an optimal fit to the experimental data.  Hydrodynamic calculations are applied after some small thermalization time, $\tau_0=0.6$ fm/$c$ in this instance. At that time, the initial correlation function $C'_{ab}(\Delta\bm{r})$ needs to be defined. Its strength is fixed by the susceptibility $\chi_{ab}$, but its initial spread is unknown. The spread along the longitudinal coordinate, Bjorken $\eta$, is assumed to be Gaussian and characterized by a width $\sigma_0$. The parameters $\sigma_0$, $T_f$ and the diffusion constant are all varied to determine the sensitivity of the various balance functions, $B_{hh'}$. For this study balance functions are studied as a function of relative rapidity and six hadronic combinations, $hh'$, are analyzed, those involve pions, kaons and protons. Four of the balance functions are compared to experimental results from STAR. Even though comparisons are premature due to the current lack of a microscopic treatment of the hadronic stage, it does appear that discrepancies between the model and data might persist, even though the model fits the data to the 10-20\% accuracy across all balance functions.

The next section provides a detailed description of the calculation. This includes the evolution of correlation functions in coordinate space, and the projection of the three-by-three correlation in coordinate space, $C'_{ab}$, onto the general charge balance functions, $B_{hh'}$ in momentum space. Sections \ref{sec:evolution} and \ref{sec:hadronization} describe the space-time evolution of $C'_{ab}$ during the hydrodynamic stage and its projection onto correlations indexed by hadronic species, $C'_{hh'}$. Section \ref{sec:results} presents results from varying the parameters mentioned above, along with a cursory comparison with results from STAR. The final section presents several sources of the discrepancies with experimental data and suggests several remedies, such as better modeling of the hadronic stage. An appendix describes the algorithm for finding the hyper-volume elements used to generate hadrons at the Cooper-Frye surface, defined by $T_f$, using the hydrodynamic history.


\section{Methods}
\label{sec:method}

Calculations are based on output from the iEBE-VISHNU relativistic hydrodynamic model \cite{Shen:2014vra}, which was performed in 2+1 dimensions, using the assumption of Bjorken boost invariance to eliminate the need to model along the $z$-axis, which is parallel to the beam. The hydrodynamic output is binned in the transverse dimensions with elements $\Delta x=\Delta y=0.1$ fm and in time with $\Delta \tau=0.02$ fm/$c$. Boost invariance assumes that matter has no longitudinal acceleration, $v_z=z/t$, and that properties are independent of the longitudinal coordinate $\eta=\sinh^{-1}(z/\tau)$. Information need only be stored for $\eta=z=0$, and the collective velocity at $\eta=0$ is purely transverse. The velocity gradient at $\eta=0$ is $dv_z/dz=1/\tau$. Here, $\tau$ is the time measured by an observer at rest relative to the longitudinal velocity, so at $\tau=t$ at $\eta=0$ and $\tau=t/\cosh\eta$ otherwise. The following information is stored for each point in the three-dimensional, $\tau,x,y$, mesh: The temperature $T$, the stress-energy tensor $T_{\alpha\beta}$ and the collective velocity $v_x$ and $v_y$. 

The equation of state was taken from lattice calculations \cite{Huovinen:2009yb}. Lookup tables were constructed so that all intrinsic quantities, such as the entropy density $s$, could be found from either the temperature or energy density. For temperatures below 155 MeV, the equation of state was assumed to be that of a hadron gas, calculated by using all resonances from the Particle Data Group with masses below 2.2 GeV \cite{pdg}. For temperatures above 175 MeV, the lattice results were used and for $155<T<175$ the equation of state was a weighted average between the two, with the weight varying between zero and unity linearly as a function of temperature in the range. The charge susceptibility matrix, $\chi_{ab}$, and the diffusion constant were taken from lattice calculations \cite{Aarts:2014nba} and stored as functions of temperature. 

The correlations were calculated Monte Carlo by generating pairs of charges. To represent the $ab$ component of a correlation the first particle in the pair was randomly, and given a charge $\pm a$. The charge of the second particle was $\pm b$ with the relative sign chosen to correspond to the sign of the source function. For example, if $S_{uu}$ is positive (creating quarks), the charges representing $C'_{uu}$ would be chosen as a $u\bar{u}$ pair or a $\bar{u}u$ pair. If one is at a point in space-time where up quarks are being annihilated, the sampling pair would be either a $uu$ or $\bar{u}\bar{u}$ pair. Once the hadronization phase is entered, one finds negative contributions to $S_{ud}$ due to the creation of pions. This results in sampling charges of either $ud$ or $\bar{u}\bar{d}$ for representing $C'_{ab}$. With this choice the charges were uncorrelated with other pairs. The decision as to whether to create a pair at a specific point in the space-time mesh was taken from the source function, Eq. (\ref{eq:source}). The number of pairs of type $ab$ to create in an element was chosen randomly with probability,
\begin{eqnarray}
\label{eq:dNfromS}
dN_{ab}&=&\tau \Delta x \Delta y\Delta \tau~|S_{ab}(x,y,\tau)|.
\end{eqnarray}
This is the number within one unit of spatial rapidity, i.e. the longitudinal effective size is $\tau\Delta\eta$. In order to avoid acausal propagation, the diffusion equation was treated as a random walk. Charges moved at the speed of light, with initial directions being random when viewed in the matter frame. At each time step, all existing charges were allowed to either re-scatter or to continue on their trajectory, with the scattering probability being $\Delta\tau/(6Du_0$), where $D$ is the diffusion constant, $\Delta\tau$ is the time step, and $u_0$ is the zeroth component of the four-velocity describing the fluid velocity. The scattering would reorient the direction so that it would again be random in the local rest frame. This probability was chosen so that the scattering rate would reproduce the desired diffusion constant for the matter according to its local temperature. Once several scatterings occur, this motion should closely mimic diffusive behavior, while for the first few scatterings, the motion avoids acausal transport by having the individual charges move at the speed of light. When choosing the lattice diffusion constant, the typical number of scatterings was a half dozen per charge.

The source in Eq.~\eqref{eq:dNfromS} was calculated using Eq. (\ref{eq:source}). If the fluid elements were to flow and expand with the fluid, so that they encompassed a fixed amount of charge, or fixed amount of entropy for ideal hydrodynamics, the change in the number of pairs within the volume $\Omega$ would be $dN_{ab}=d(\chi_{ab}\Omega)$, the change of the charge susceptibility over the volume of the fluid element. Because the fluid elements had fixed transverse sizes, $\Delta x$ and $\Delta y$, a longitudinal size of $\tau \Delta \eta$ (with $\Delta \eta=1$ unit of spatial rapidity), and were evaluated in a time interval $\Delta \tau$, the number of pairs created in the element  with coordinates $x',y',\eta',\tau'$ such that $x<x'<x+\Delta x,y<y'<y+\Delta y,0<\eta'<1,\tau<\tau'<\tau+\Delta \tau$ is
\begin{eqnarray}
\label{eq:dNab}
dN_{ab}&=&\int_{r'\in \Delta^4r} d^4r'~[\partial_t\chi_{ab}(r')
+\bm{\nabla}(\chi_{ab}(r')\bm{v}(r'))]\\
\nonumber
&=&\int d^4r'~\partial_\mu(\chi_{ab}u^\mu)=\oint d\Omega_\mu~u^\mu\chi_{ab}\\
\nonumber
&=&\frac{\Delta x~\Delta y}{4}\left\{
(\tau+\Delta \tau)u_0(x+\Delta x,y+\Delta y,\tau+\Delta \tau)\chi_{ab}(x+\Delta x,y+\Delta y,\tau+\Delta \tau)\right.\\
\nonumber
&+&(\tau+\Delta \tau)u_0(x+\Delta x,y,\tau+\Delta \tau)\chi_{ab}(x+\Delta x,y,\tau+\Delta \tau)\\
\nonumber
&+&(\tau+\Delta \tau)u_0(x,y+\Delta y,\tau+\Delta \tau)\chi_{ab}(x,y+\Delta y,\tau+\Delta \tau)\\
\nonumber
&+&(\tau+\Delta \tau)u_0(x,y,\tau+\Delta \tau)\chi_{ab}(x,y,\tau+\Delta \tau)\\
\nonumber
&-&\tau u_0(x+\Delta x,y+\Delta y,\tau)\chi_{ab}(x+\Delta x,y+\Delta y,\tau)
-\tau u_0(x+\Delta x,y,\tau)\chi_{ab}(x+\Delta x,y,\tau)\\
\nonumber
&-&\tau u_0(x,y+\Delta y,\tau)\chi_{ab}(x,y+\Delta y,\tau)
-\left.\tau u_0(x,y,\tau)\chi_{ab}(x,y,\tau)\right\}\\
\nonumber
&+&\frac{\Delta y~\Delta \tau}{4}\left\{
(\tau+\Delta \tau)u_x(x+\Delta x,y+\Delta y,\tau+\Delta \tau)\chi_{ab}(x+\Delta x,y+\Delta y,\tau+\Delta \tau)\right.\\
\nonumber
&+&(\tau+\Delta \tau)u_x(x+\Delta x,y,\tau+\Delta \tau)\chi_{ab}(x+\Delta x,y,\tau+\Delta \tau)\\
\nonumber
&-&(\tau+\Delta \tau)u_x(x,y+\Delta y,\tau+\Delta \tau)\chi_{ab}(x,y+\Delta y,\tau+\Delta \tau)\\
\nonumber
&-&(\tau+\Delta \tau)u_x(x,y,\tau+\Delta \tau)\chi_{ab}(x,y,\tau+\Delta \tau)\\
\nonumber
&-&\tau u_x(x+\Delta x,y+\Delta y,\tau)\chi_{ab}(x+\Delta x,y+\Delta y,\tau)
-\tau u_x(x+\Delta x,y,\tau)\chi_{ab}(x+\Delta x,y,\tau)\\
\nonumber
&+&\tau u_x(x,y+\Delta y,\tau)\chi_{ab}(x,y+\Delta y,\tau)
+\left.\tau u_x(x,y,\tau)\chi_{ab}(x,y,\tau)\right\}\\
\nonumber
&+&\frac{\Delta x~\Delta \tau}{4}\left\{
(\tau+\Delta \tau)u_y(x+\Delta x,y+\Delta y,\tau+\Delta \tau)\chi_{ab}(x+\Delta x,y+\Delta y,\tau+\Delta \tau)\right.\\
\nonumber
&-&(\tau+\Delta \tau)u_y(x+\Delta x,y,\tau+\Delta \tau)\chi_{ab}(x+\Delta x,y,\tau+\Delta \tau)\\
\nonumber
&+&(\tau+\Delta \tau)u_y(x,y+\Delta y,\tau+\Delta \tau)\chi_{ab}(x,y+\Delta y,\tau+\Delta \tau)\\
\nonumber
&-&(\tau+\Delta \tau)u_y(x,y,\tau+\Delta \tau)\chi_{ab}(x,y,\tau+\Delta \tau)\\
\nonumber
&+&\tau u_y(x+\Delta x,y+\Delta y,\tau)\chi_{ab}(x+\Delta x,y+\Delta y,\tau)
-\tau u_y(x+\Delta x,y,\tau)\chi_{ab}(x+\Delta x,y,\tau)\\
\nonumber
&+&\tau u_y(x,y+\Delta y,\tau)\chi_{ab}(x,y+\Delta y,\tau)
-\left.\tau u_y(x,y,\tau)\chi_{ab}(x,y,\tau)\right\}.
\end{eqnarray}

At the initial time $\tau_0$, the susceptibility is assumed to have already reached its equilibrium value, which necessitates that a number of pairs already exist. For the points $x',y',z',\tau'=\tau_0$, the number of pairs created is
\begin{eqnarray}
\label{eq:dNab0}
dN_{ab}&=&\frac{\Delta x~\Delta y}{4}\tau_0\left\{\chi_{ab}(x+\Delta x,y+\Delta y,\tau_0)
+\chi_{ab}(x+\Delta x,y,\tau_0)\right.\\
\nonumber
&+&\left.\chi_{ab}(x,y+\Delta y,\tau_0)
+\chi_{ab}(x,y,\tau_0)\right\}.
\end{eqnarray}
For the current run, there was no initial transverse flow, otherwise more terms would be added as in Eq. (\ref{eq:dNab}). The two particles from one of these initial pairs were initially placed at the transverse coordinate $\bm{x}_\perp = (x,y)$, according to Eq. (\ref{eq:dNab0}). However, the two partners were not placed at the same longitudinal coordinate. Instead, each charge was randomly moved from a given starting point by a Gaussian random number with standard deviation $\sigma_0$, effectively stating that by the time the hydrodynamic treatment was started, each existing charge would have diffused in spatial rapidity from the point at which that charge and its partner had been created. Physically, this initial separation might have little to do with diffusive dynamics, and instead might be due to the process of charges being created pair-wise by breaking strings or in the decay of longitudinal chromo-electric fields. To convert the field, or string, energy into particles, the charges might tunnel a certain distance on the order of a unit of rapidity. Here, $\sigma_0$ is treated as an unknown parameter. If $\sigma_0$ is much less than the subsequent diffusion distance, $\sigma_0$ becomes irrelevant. Given that the diffusion distance is expected to be of the order of a unit of rapidity, results should be sensitive to values of $\sigma_0$ more than a few tenths of spatial rapidity. The initial separation of the two charges from the same pair in the same transverse direction was neglected. The initial longitudinal separation is critical because the large initial velocity gradient, $dv_z/dz=1/\tau$, magnifies any small initial spread. Because there is little initial transverse velocity gradient, small initial transverse separations should not significantly affect the final separations of the charges.

The sample charges eventually cross the hyper-surface and are translated into hadrons. For each of these small charges, $\delta Q_a$, a small change is seen in the number of emitted hadrons of a specific species $\delta N_h$. The hyper-surface element $\Omega_\mu$ is characterized by its direction, which is proportional to $-\partial_\mu T$, and the collective four-velocity at that point, $u^\mu$. The Cooper-Frye formula is adjusted by modifying the phase space density for hadron species $h$ according to a small chemical potential, $\delta \mu_a/T=-\delta \alpha_a$,
\begin{eqnarray}
f_h(\bm{p},\bm{r},t)\rightarrow f_h(\bm{p},\bm{r},t)e^{-\delta\alpha_a q_{h,a}},
\end{eqnarray}
where $q_{h,a}$ is the charge of type $a$ carried by a single hadron of type $h$. The additional net charge induced by the chemical potential must reproduce the small charge $Q_a$ that passes through the hyper-surface element $\Omega_\mu$,
\begin{eqnarray}
\label{eq:Mabint}
\delta Q_a&=&-\Omega_\mu\sum_h\int \frac{d^3p}{(2\pi)^3E_{hp}}\frac{p^\mu}{q_{ha}}f_h(\bm{q},\bm{r},t)\Theta(p\cdot\Omega)q_{hb} \delta \alpha_b,\\
\nonumber
&=&M_{ab} \delta\alpha_b,\\
\nonumber
M_{ab}&\equiv&-\Omega_\mu\sum_h \frac{d^3p}{(2\pi)^3E_{hp}}\frac{p^\mu}{q_{ha}}f_h(\bm{q},\bm{r},t)\Theta(p\cdot\Omega) q_{hb}.
\end{eqnarray}
The choice of whether to include the step function $\Theta(p\cdot\Omega)$ is somewhat arbitrary. The step function excludes emission into the surface, which occurs when the hyper-surface is receding into the matter at subluminal speeds. In matter that speed is
\begin{eqnarray}
v_\Omega&=&\frac{d\Omega_0}{|d\bm{\Omega}|}.
\end{eqnarray}
For $v_\Omega>1$ the step function is irrelevant. For elements where $v_\Omega<1$, keeping the step function eliminates emission into the surface but also leads to a violation of energy and momentum conservation. Studies of heavy ion collisions show that this backflow contribution is less than one percent of the total emission \cite{Pratt:2010jt}, so for calculating $M_{ab}$, it is neglected here \cite{Pratt:2014vja}. Once the step function is eliminated from the equation, parity eliminates the part of the integral proportional to $\bm{p}\cdot \bm{\Omega}$ when calculated in the matter frame, and $M_{ab}$ and the number of hadrons $N_h$ emitted normally through the surface is
\begin{eqnarray}
\label{eq:Mabdef}
M_{ab}&=&(u\cdot\Omega)\chi_{ab},\\
\nonumber
\chi_{ab}&=&\sum_h n_hq_{ha}q_{hb},\\
\nonumber
N_h&=&(u\cdot\Omega)n_h,
\end{eqnarray}
where $n_h$ is the density of hadrons in an equilibrated system of type $h$ at the interface temperature. One can now solve for the chemical potential and find the additional emission,
\begin{eqnarray}
\label{eq:alphaa}
\delta\alpha_a&=&M^{-1}_{ab}\delta Q_b,\\
\nonumber
\delta N_{h}&=&n_h \chi^{-1}_{ab}q_{ha}\delta Q_b,
\end{eqnarray}
and the number of additional emitted particles of type $h$, $\delta N_h$, depends only on the additional charge that flows through the surface, $\delta Q_a$, and not on the size of the surface element. Hadrons were produced randomly via Monte Carlo so that the number of hadrons $\delta N_h$ would be reproduced on average. Momenta were assigned consistent with the interface temperature, thermal motion and viscous corrections according to the methods described in \cite{Pratt:2010jt}. Unstable particles were decayed statistically according to weights taken from the Particle Data Group \cite{pdg}.

The numerator,
\begin{eqnarray}
\label{eq:numdef}
\mathcal{N}_{hh'}(\Delta y)&=&\int dydy'~\langle (N_h(y)-N_{\bar h}(y))(N_{\bar{h}'}(y')-N_{h'}(y')\rangle\delta(\Delta y-|y-y'|),
\end{eqnarray}
for the charge balance function is then constructed by binning on the measure of relative momentum, e.g. $\Delta y$, whenever two particles of type $h$ or $h'$ are found with $h$ and $h'$ resulting from the charge pairs of type $a$ and $b$ respectively from the source function at $S_{ab}d^4x$. Particles from charges that originated from different points in space-time are uncorrelated because the charges of type $a$ were randomly assigned to positive or negative, and the charges of the type-$b$ charges were picked to be the same or opposite as the $a$ particles from the same space-time point according to the sign of $S_{ab}$. Thus, by attaching information to identify particles from the same space-time point, only those pairs from the same space-time point were used to construct the numerator, which greatly reduces the noise inherent in the calculation and allows smooth results to be generated with only a few minutes of CPU time. Also, to increase the efficiency of the calculation, charges were created with an oversampling factor of $M_c$ and hadrons for a given charge were created with an over-sampling factor of $M_h$. The numerator and denominators of the balance function were then scaled to account for the over-sampling.

Decays also represent additional sources of correlation. For example a neutral particle might decay into two charged particles which must then be included in the balance function numerator. This is accounted for by producing an ensemble of uncorrelated particles from the hyper-surface, unrelated to the sampling charges. This sampling ignores charge conservation, as each emission is independent. However, the decays of such resonances do lead to additional correlations. Particles from the same decay are then evaluated for their contribution to the binning of the numerator $\mathcal{N}_{hh'}(\Delta y)$. The ensembles created here are the same ones that are used to calculate the denominator of the balance function described below.

For a charge neutral system, the charge balance functions from Eq. (\ref{eq:baldef}) can be equivalently expressed in terms of $\mathcal{N}_{hh'}$,
\begin{eqnarray}
\label{eq:bfdef}
B_{hh'}(\Delta y)&=&\frac{\mathcal{N}_{hh'}(\Delta y)}{\langle N_h+N_{\bar{h}}\rangle},
\end{eqnarray}
The choice to define the balance function with a division by the average number is motivated by having a result that integrates to some constant that is quasi-independent of multiplicity. If $h$ and $h'$ both refer to all positive particles (and $\bar{h}$ and $\bar{h}'$ refer to all negative particles), the balance function would integrate to unity for perfect acceptance. If the system were not charge-neutral, then the definition in Eq. (\ref{eq:baldef}) would still integrate to unity, and for that reason Eq. (\ref{eq:baldef}) would be a more attractive option for a system with a significant net charge, such as what would be encountered in the beam-energy scan program planned for RHIC in the coming years. 

Calculating the denominator, $\langle N_h+N_{h'}\rangle$ in Eq. (\ref{eq:bfdef}), does not involve the charges that represent the numerator. Instead, it relies only on the list of hyper-surface elements and their properties.  The number of such particles is for a hyper-surface element $\Omega$ is
\begin{eqnarray}
N_{h}&=&(2S_h+1)\int\frac{d^3p}{(2\pi\hbar)^3E}(p\cdot\Omega)f_h(p). 
\end{eqnarray}
Again, because the step function, as in Eq. (\ref{eq:Mabint}), is ignored, the integral is easy to calculate in the frame of the fluid. In that frame the zeroth component of $\Omega$ is $u\cdot\Omega$, and $N_h=n_hu\cdot\Omega$, as mentioned in Eq (\ref{eq:Mabdef}), where $n_h$ is the density of hadrons. Hadrons are generated Monte Carlo consistent with $N_h$. The momentum spectrum of the particles is corrected for shear anisotropy using the methods of \cite{Pratt:2010jt}. Also arbitrarily, if the momentum of a particle was found to be emitted into the surface, i.e. $p\cdot\Omega<0$, the particle was regenerated. Once emitted, the unstable particles are decayed. 

For $h$ and $h'$ referring to all charged particles, the balance function should integrate to unity. This tests the accuracy of the procedure. Given that the balance-function numerator is calculated using the representative sample charges to generate hadrons as they pass through the hyper-surfaces, and that the denominator uses only the hyper-surface elements, the test is highly non-trivial. For the calculations presented here, the balance function normalization came within one or two tenths of a percent of unity.

Experimental acceptance plays a pivotal role in the balance function. For example, the normalization of the all-charge balance function is reduced to approximately 0.3. This means that for each observed charge, the number of additional observed opposite charges, vs the number of additional same-sign charges was approximately thirty percent. The other seventy percent of the balancing charge, was either outside the acceptance, or not recorded due to the efficiency which is typically about 70\% for identified particles. If one used those charges that were not identified as pions, kaons or protons, the efficiency would be somewhat higher.

Binning the balance function is rather straight-forward. One performs a sum of all hadron pairs where one came from the corresponding $ab$ elements of the source function. If the particles were charged pions, charged kaons, protons, or antiprotons, the corresponding bin in relative rapidity was incremented. For the denominator, a sum over all hyper elements was evaluated, with hadrons being created and decayed as mentioned above. For calculating the denominator, hadrons were boosted so that their rapidities would randomly cover a region $-\eta_{\rm max}<y<\eta_{\rm max}$, with $\eta_{\rm max}=0.9$ being large enough that no particles with that rapidity could be detected. For evaluating the numerator, the particle generated by charge $a$ was also boosted randomly, and the particle generated by charge $b$ was boosted by the same amount so that the relative rapidity was unchanged.


\section{Hydrodynamic Evolution of the Correlation Function}
\label{sec:evolution}

\begin{figure}
\centerline{\includegraphics[width=0.45\textwidth]{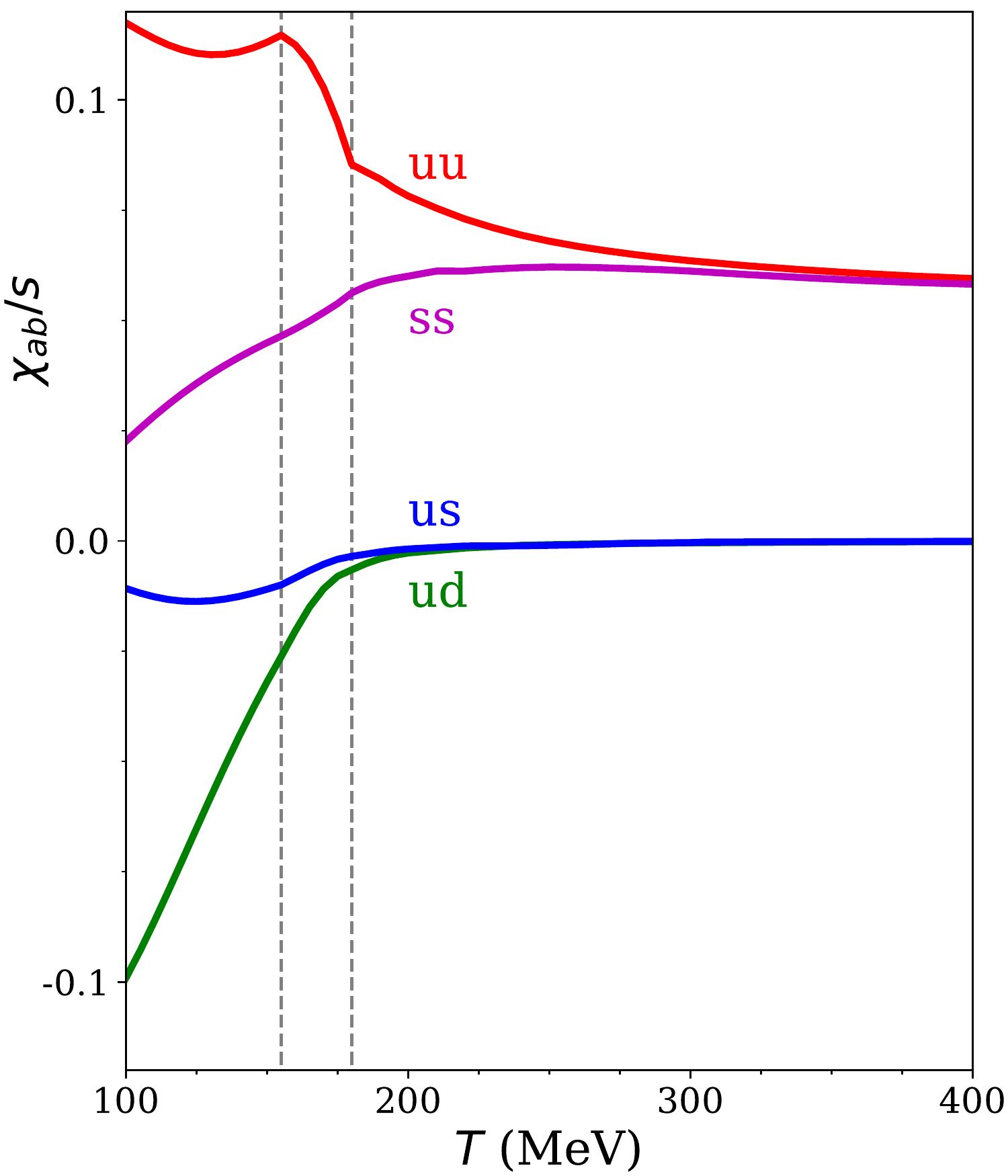}}
\caption{\label{fig:chilattice}(color online)
The off-diagonal components of $\chi_{ab}$ vanish at high temperature as quarks evolve independently, but at low temperatures the combination of multiple quarks within a single hadron drives strong off-diagonal elements. For a fixed amount of entropy, an expanding volume element strongly increases the amount of charge as it reaches the hadronization region, again due to the fact that multiple charges occupy an individual hadron, even though the entropy per hadron in the hadron phase is similar to the entropy per quark in the plasma phase. For regions to the right of the dashed lines, lattice calculations were used, while left of the dashed lines displays the results for a hadron gas. The intermediate region used an interpolation between the two calculations, $155 < T< 175$ MeV.}
\end{figure}

\begin{figure}
\centerline{\includegraphics[width=0.65\textwidth]{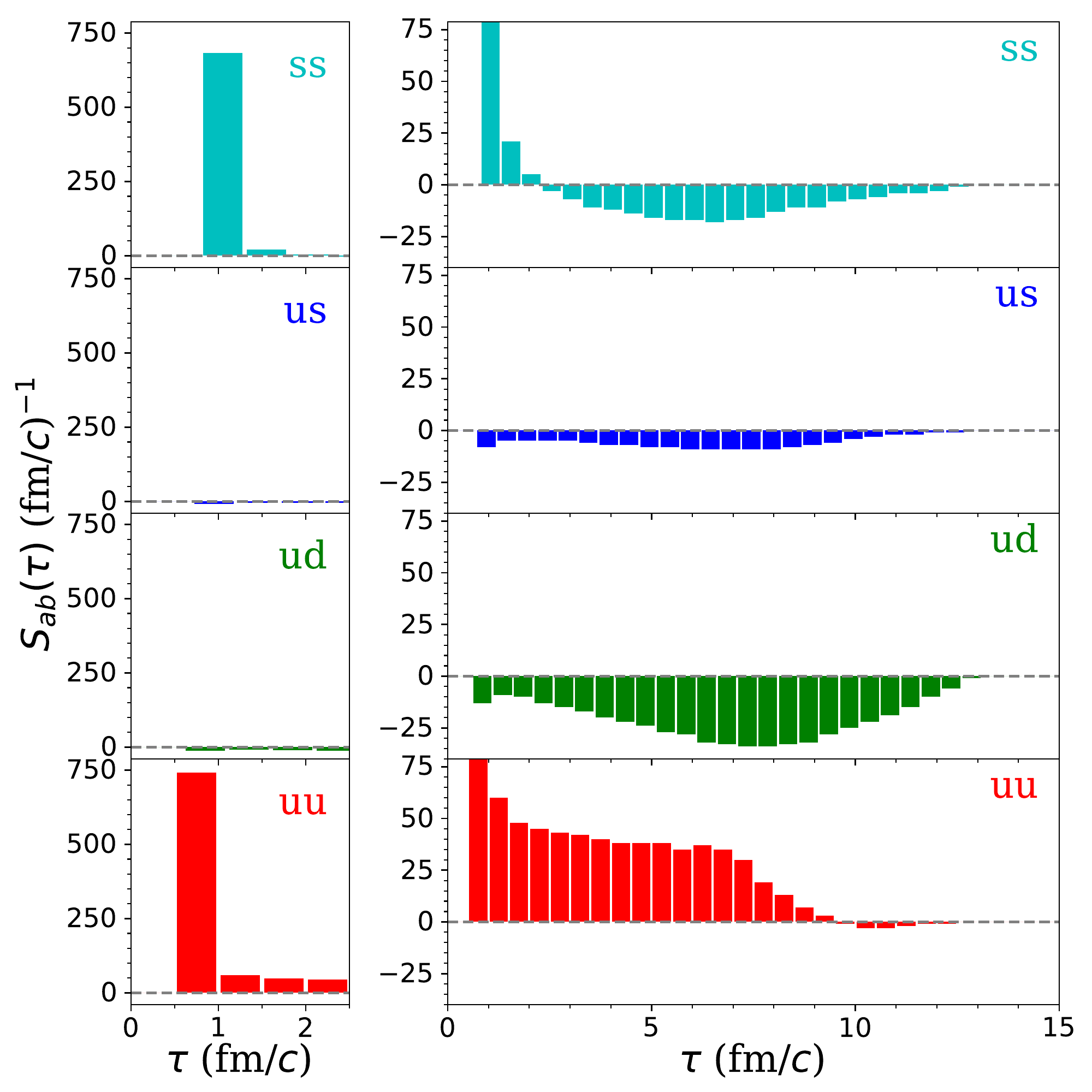}}
\caption{\label{fig:udssource}(color online)
The source function that generates the correlation function $C'_{ab}(\Delta\eta)$ is shown as a function of proper time $\tau$. The function is found by convoluting the susceptibility from lattice calculations with the hydrodynamic evolution as prescribed in Eq. (\ref{eq:source2}). Multiplied by the bin width, 0.5 fm/$c$, the values give the number of produced charge pairs within the time step that sample the evolution. The initial spike of the source function describes the correlation created at initial thermalization, $\tau_0$. }
\end{figure}

As prescribed by Eq. (\ref{eq:source}), the source function that drives the growth of the correlation function, $C'_{ab}(\Delta \eta)$, is prescribed by the rate of change of the susceptibility, shown as a function of temperature in Fig. \ref{fig:chilattice}. For the correlation function in relative rapidity and for a boost-invariant system, integrating Eq. (\ref{eq:source}) over transverse coordinates becomes
\begin{eqnarray}
\label{eq:source2}
\frac{\partial}{\partial\tau}C'_{ab}(\Delta\eta)&=&S_{ab}(\tau)=
\frac{\partial}{\partial\tau}\left\{\tau \int dxdy~\chi_{ab}(x,y,\tau)\gamma_\perp(x,y,\tau)\right\},
\end{eqnarray}
where $\gamma_\perp$ is the Lorentz contraction factor for the transverse flow, $\gamma_\perp=\sqrt{1+u_x^2+u_y^2}$. In a time $d\tau$ the number of created sampling charge pairs of type $ab$ within one unit of rapidity is $S_{ab}(\tau)d\tau$.  The sampling charges are created at each space-time point according to Eq. (\ref{eq:dNab}). The charges are tagged so that they can be paired with the partner charge from the same pair at the end of the collisions. The susceptibilities from lattice as a function of temperature are shown in Fig. \ref{fig:chilattice}, and the resulting source functions are displayed in Fig. \ref{fig:udssource}, where the hydrodynamic evolution was taken from descriptions of full-energy Au+Au collisions at RHIC. Assuming isospin symmetry, $\chi_{uu}=\chi_{dd}$, $\chi_{us}=\chi_{ds}$, and $S_{uu}=S_{dd}$ and $S_{us}=S_{ds}$, there are four independent components of $\chi$ and of $S$: $\chi_{uu}$, $\chi_{ss}$, $\chi_{ud}$ and $\chi_{us}$, and $S_{uu}$, $S_{ss}$, $S_{ud}$ and $S_{us}$.

As described in the introduction, susceptibilities provide valuable insight into the quark chemistry at a given temperature. At high temperatures, quarks are uncorrelated with one another, and the only correlation comes from a quark with itself (plus a small correction due to the Pauli-exclusion principle). Because the density of quarks for a massless quark parton gas is proportional to the entropy density, one expects $\chi_{uu}/s\approx\chi_{dd}/s\approx\chi_{ss}/s$ at high temperature and that those values should vary little with $T$. It is remarkable to see how $\chi_{ab}/s$ from the lattice calculations in Fig. \ref{fig:chilattice} approach this Stefan-Boltzmann limit for temperatures only modestly higher than 200 MeV. In contrast, at low temperatures quarks form hadrons, and even though hadrons are independent of one another, there exist correlations between quarks within a hadron. For example for a $\Delta^{++}$ baryon ($uuu$) contributes a factor of 9 to the $\chi_{uu}$ and a $K^+$ meson ($u\bar{s}$) contributes +1 to $\chi_{uu}$, +1 to $\chi_{ss}$, and -1 to $\chi_{us}$. Because of the heavier mass of strange hadrons, $\chi_{ss}/s$ falls with temperature as $T$ falls in the hadron region. In contrast, $\chi_{uu}$ rises due to the fact that a large number of up and down quarks must appear as the system goes through hadronization. If one were to plot the hadron susceptibility,
\begin{eqnarray}
\chi_{BB}/s&=&\frac{1}{9s}\left(4\chi_{uu}+\chi_{dd}+\chi_{ss}+4\chi_{ud}+4\chi_{us}+2\chi_{ds}\right)\\
\nonumber
&\approx&\frac{1}{9s}\left(5\chi_{uu}+\chi_{ss}+4\chi_{ud}+6\chi_{us}\right),
\end{eqnarray}
one would see that the baryon susceptibility would fall precipitously for low $T$ due to the large thermal penalty coming from the heavy baryon masses.

It is assumed that chemical equilibrium has been attained by the time the hydrodynamic evolution has begun. Thus, a large number of pairs are created in the first instant, which gives a peak to $S_{ab}(\tau\approx \tau_0=0.6~{\rm fm}/c)$. This initial surge, along with the continuous contribution for later times is displayed in Fig. \ref{fig:udssource}. This was generated from the hydrodynamic solution for zero-impact-parameter collisions of $Au$ nuclei at the highest RHIC energy, $\sqrt{s_{NN}}=200$ GeV.   Unlike the correlation at later times, the initial correlation does not increment $C'_{ab}(\Delta\eta)$ at $\Delta\eta=0$. Instead, both charges are spread randomly in spatial rapidity with a Gaussian distribution described by $\sigma_0$, to account for the fact that the correlation could have been sourced at any time $\tau<\tau_0$, and thus may have already spread significantly. The sensitivity of the final observed correlations to $\sigma_0$ is investigated in Sec. (\ref{sec:results}). 

\begin{figure}
\centerline{\includegraphics[width=0.45\textwidth]{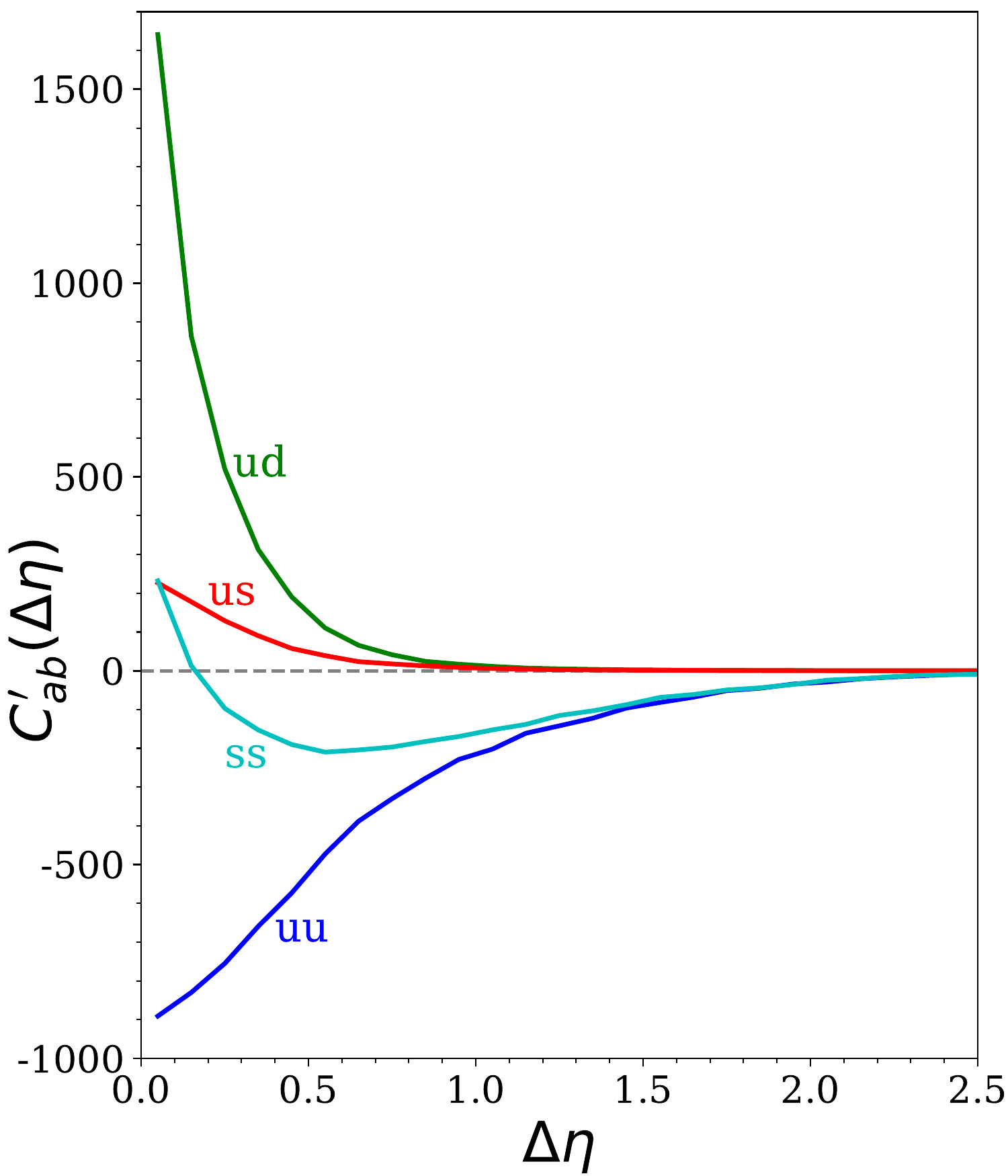}}
\caption{\label{fig:cf_uds}(color online)
The correlation $C'_{ab}$ as a function of relative rapidity at breakup. The $uu$, $dd$ and $ss$ components have an extended range because they originate from the original creation of charge at $\tau_0$. The off-diagonal elements, $ud$, $ud$ and $ds$ arise from the creation of mesons at hadronization. The reduction of strangeness as the system falls below $T_c$ explains the rise in $C'_{ss}$ at small $\Delta\eta$. 
}
\end{figure}

The charges propagate with straight trajectories at the speed of light, punctuated with randomizing collisions as described in Sec. \ref{sec:method}. Each charge ultimately crosses the hyper-surface defining the boundary with the hadronic phase. At this point, each pair of charges was binned according to the relative spatial rapidity, $\Delta\eta$, with its partner. This produces a statistical sampling of the correlation function, $C'_{ab}(\Delta\eta)$, in relative spatial rapidity at decoupling, and is displayed in Fig. \ref{fig:cf_uds}. Comparing with the source function, $S_{ab}(\tau)$, in Fig. \ref{fig:udssource}, one can see how contributions from $S_{ab}(\tau)$ for early times leads to broader peaks in the correlation function. This is especially true for the initial correlation generated by $S_{ab}(\tau_0)$. Not only does this correlation have more time to spread, it may have been created with a significant width if $\sigma_0$ is large. Given that all correlations are ultimately measured in momentum space, and are thus spread out by the thermal motion, having $\sigma_0\gtrsim 0.5$ should lead to a significant broadening of the correlation derived from the initial source. 

The $uu$ component of the source is strong and indeed generates long range structure to $C_{uu}$ from the initial correlation. Similarly, the $ss$ component from the initial correlation translates into a long-range component for the strange quarks. Because charge balance enhances charges of the opposite sign, these long-range contributions are negative. Due to the surge of new up and down quarks that appear during hadronization, the ensuing correlation for $uu$ charge has additional negative contributions that then contribute to $C_{uu}$ for small $\Delta\eta$. In contrast, the number of strange quarks can fall at later times, and this contribution becomes negative for $C_{ss}$, so that $C_{ss}$ has a significant positive contribution at later times. The off-diagonal contributions, $ud$ and $us$, are non-existent for a plasma of independent quarks, and the corresponding source functions come in at later times. Given the much larger number of up and down, relative to strange quarks, in the hadron phase, the correlation for $ud$ is much stronger than that for $us$ in Fig. \ref{fig:cf_uds}. 

.


\section{Hadronization}
\label{sec:hadronization}

Unfortunately, measurements of $C'_{ab}$ are not currently possible. The non-measurement of neutrons, plus the violation of strangeness in weak decays makes this difficult. Fortunately, three species of hadrons can be measured and identified with high efficiency: pions, kaons, protons, and their antiparticles. For correlations of the type, $C'_{hh'}$, this leads to six correlation functions: $C'_{\pi\pi}$, $C'_{\pi K}$, $C'_{\pi p}$, $C'_{KK}$, $C'_{Kp}$ ,$C'_{pp}$. The $u,d$ and $s$ charges sampling $C'_{ab}$, eventually cross the hyper-surface into the hadronic realm. The linear response described in Eq. (\ref{eq:alphaa}) translates a differential charge, $\delta Q_a$, passing through the hyper-surface into a differential number of hadrons, $\delta N_h$. This can be applied to derive $C'_{hh'}$ from $C'_{ab}$.
\begin{eqnarray}
\label{eq:cfh}
C'_{hh'}(\Delta \eta)&=&\int d\eta d\eta'~\langle(N_h(\eta)-N_{\bar{h}}(\eta))(N_{\bar{h}'}(\eta')-N_{h'}(\eta'))\rangle \delta(\Delta\eta-|\eta-\eta'|)\\
\nonumber
&=&\sum_{ab}K_{hh';ab} C'_{ab}(\Delta\eta),\\
\nonumber
K_{hh';ab}&=&-4n_hn_{h'}q_{hc}q'_{h'd}\chi^{-1}_{ca}\chi^{-1}_{db}.
\end{eqnarray}
The negative sign in the definition of the kernel comes from having defined $C_{hh'}$ as positive when there are positive correlations between hadrons of type $h$ with antiparticles of type $\bar{h}'$.

\begin{figure}
\centerline{\includegraphics[width=0.4\textwidth]{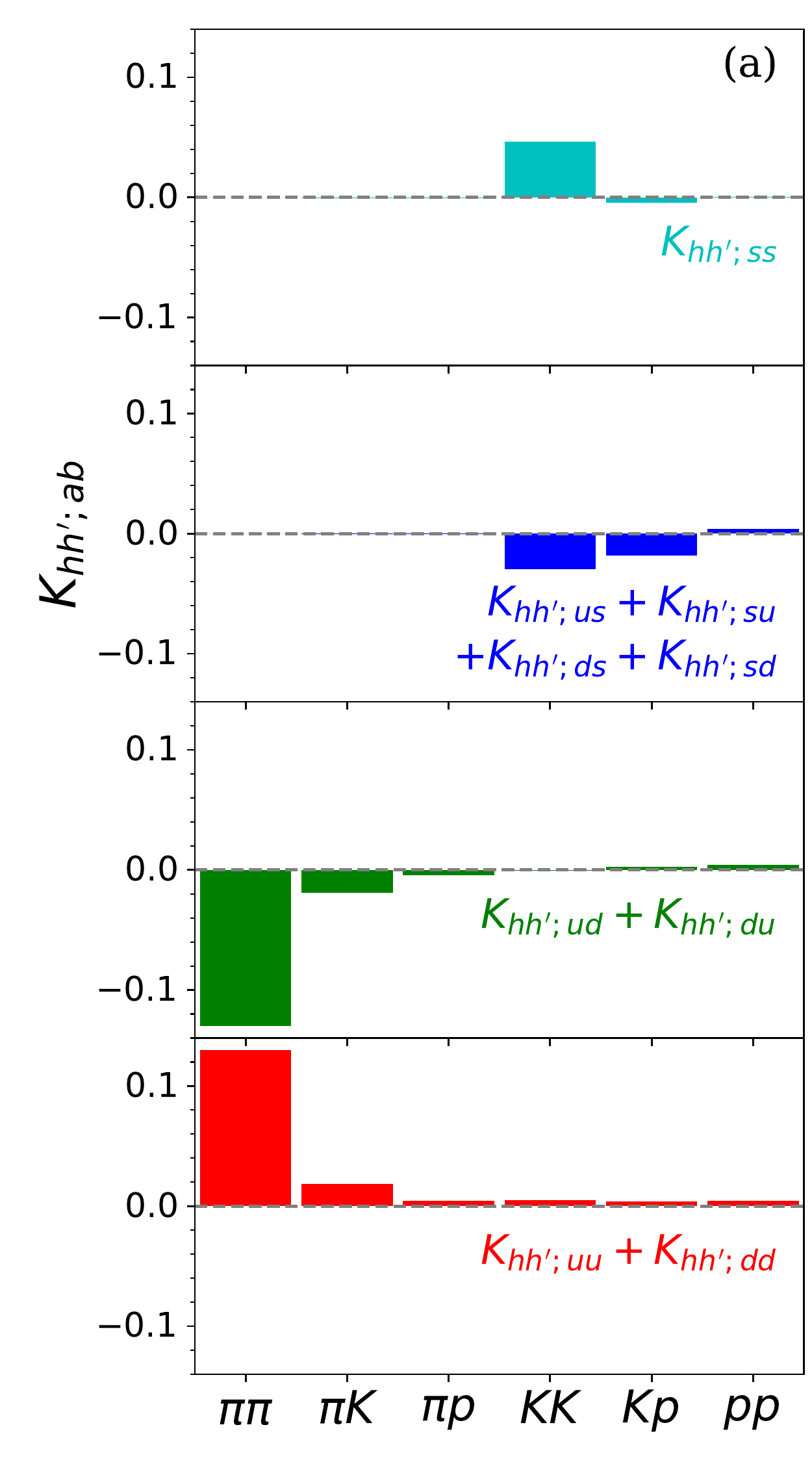}
\vspace*{0.05\textwidth}
\includegraphics[width=0.4\textwidth]{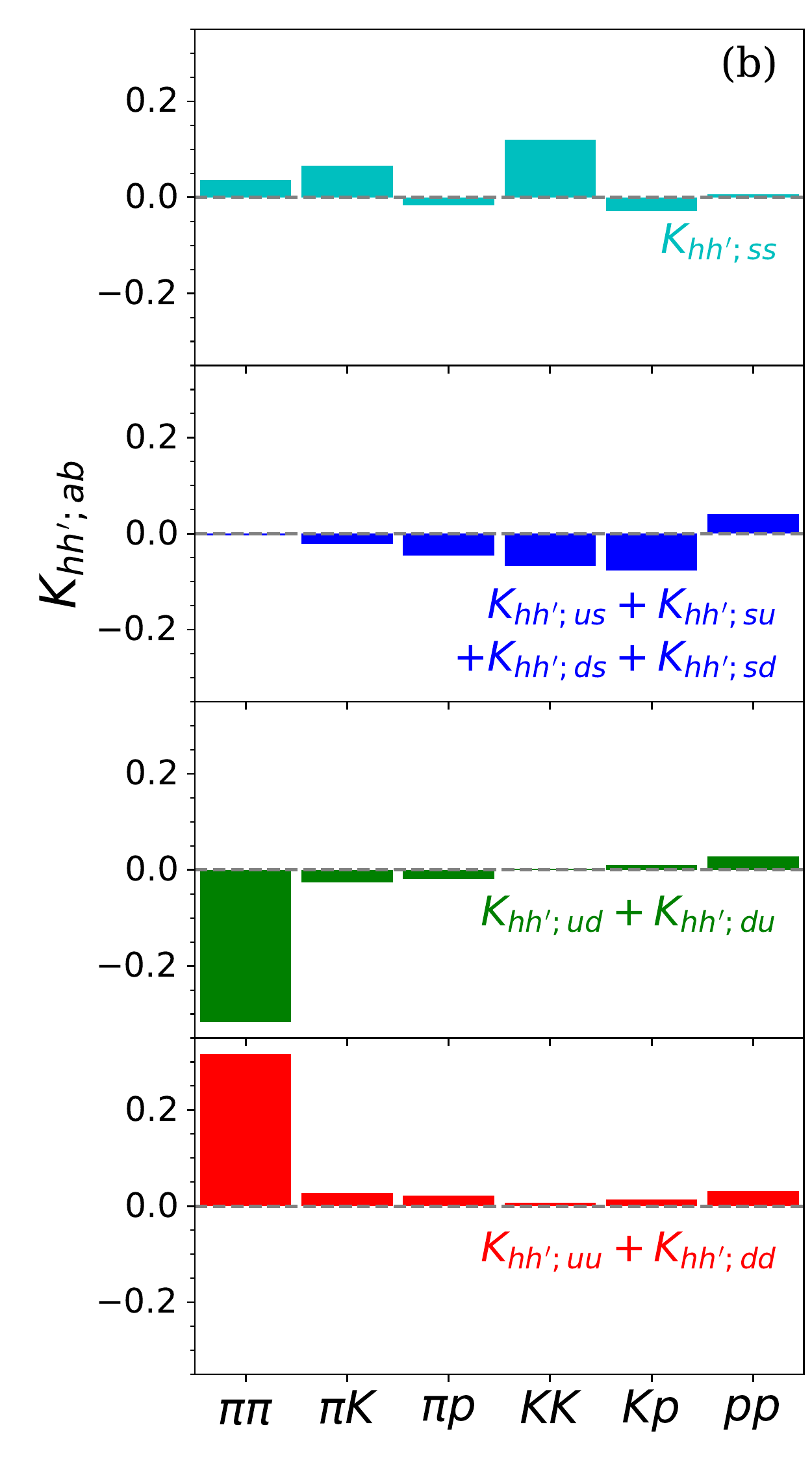}
}
\caption{\label{fig:chitrans}(color online)
The kernel $K_{hh';ab}$ translates correlations indexed by charge, $a,b=u,d,s$, to those indexed by hadron species, $h,h'$=$\pi,K,p$. After convoluting with the susceptibility $\chi_{ab}$ or the source function $S_{ab}$, due to the isospin symmetry of $\chi$ and $S$, only four independent combinations of $K_{hh';ab}$ contribute, and are shown above. The right-side panel (b) differs from (a) by including the effects of decays as described in Eq. (\ref{eq:feeddown}). Feed down decays are especially important for spreading the effect of correlations involving strange quarks amongst all hadron species. These values assumed an interface temperature of 155 MeV.}
\end{figure}
The kernel $K_{hh';ab}$ can be represented as a $6\times 4$ matrix. Using isospin symmetry between the $u$ and $d$ indices, $C'_{ab}$ has 4 independent combinations of $ab$ that contribute.
\begin{eqnarray}
C'_{hh'}(\Delta\eta)&=&\sum_{\alpha=1}^4 \tilde{K}_{hh';\alpha}\tilde{C}'_{\alpha}(\Delta\eta),\\
\nonumber
\tilde{K}_{hh';1}&=&K_{hh';uu}+K_{hh';dd},~\tilde{C}'_1(\Delta\eta)=C'_{uu}(\Delta\eta)=C'_{dd}(\Delta\eta),\\
\nonumber
\tilde{K}_{hh';2}&=&K_{hh';ud}+K_{hh';du},~\tilde{C}'_2(\Delta\eta)=C'_{ud}(\Delta\eta)=C'_{du}(\Delta\eta),\\
\nonumber
\tilde{K}_{hh';3}&=&K_{hh';us}+K_{hh';su}+K_{hh';ds}+K_{hh';sd},~\tilde{C}'_3(\Delta\eta)=C'_{us}(\Delta\eta)=C'_{su}(\Delta\eta)
=C'_{ds}(\Delta\eta)=C'_{sd}(\Delta\eta),\\
\nonumber
\tilde{K}_{hh';4}&=&K_{hh';ss},~\tilde{C}'_4(\Delta\eta)=C'_{ss}(\Delta\eta).
\end{eqnarray}
For example, even though $K_{\pi K;us}$ is significant, $K_{\pi K;ds}$ is the nearly the same but with opposite sign, thus they cancel one another when calculating $C'_{hh'}(\Delta\eta)$.

Because pions, kaons and protons are the best measured outgoing hadrons, we presently consider only six combinations of $hh'$: $\pi\pi$, $\pi K$, $\pi p$, $KK$, $Kp$ and $pp$. At this point balance functions the STAR Collaboration at RHIC has measured  four of the six hadron combinations, $\pi\pi$, $KK$, $Kp$ and $pp$. Figure \ref{fig:chitrans} shows these kernels for a transition temperature of $T=155$ MeV. The elements are stronger when the charge indices match the quark composition of the hadrons. For instance $K_{\pi\pi;ud}$ and $K_{\pi\pi;uu}$ are both large. From the kernel one can see that the $Kp$ balance function will be driven largely by the $us$ and $ss$ correlations. Because $ss$ and $us$ correlations have opposite sign, the two contributions to the balance functions have opposite sign. Further, one can see from Fig. \ref{fig:udssource} that the $ss$ correlation is established very early in the collision when the quarks are initially produced, whereas the $us$ correlation is seeded late, when kaons begin to appear. Thus, the $Kp$ balance function has a positive contribution, which will be narrow in relative rapidity, and a negative contribution that extends further in rapidity. The weight between these two contributions provides sensitive insight into the evolution of the charge-charge correlation.

\begin{figure}
\centerline{\includegraphics[width=0.48\textwidth]{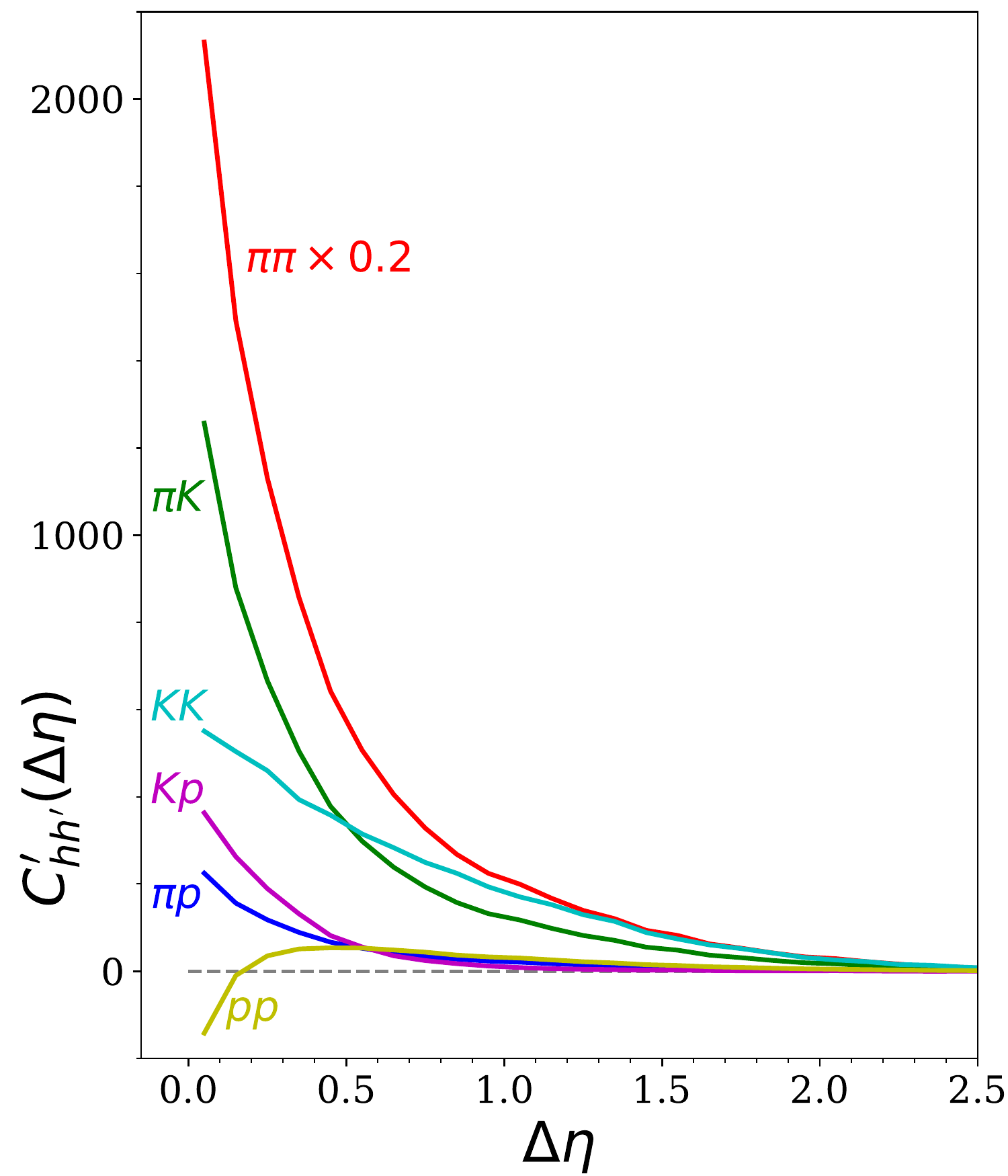}}
\caption{\label{fig:cfh}(color online)
Hadronic correlations in coordinate space immediately after hadronization.}
\end{figure}
Figure \ref{fig:cfh} shows the correlation function $C'_{hh'}(\Delta\eta)$ calculated from Eq. (\ref{eq:cfh}). As expected, $C'_{\pi\pi}$ is approximately an order of magnitude stronger than the other correlations at small $\Delta\eta$. The $KK$ correlations are broad because they are dominated by the $ss$ correlations, which are seeded at early times. The $pp$ correlations are not only seeded early, but have negative contributions to the source function at later times. This leads to the dip in the correlation function at small $\Delta\eta$. Due to the factors of density in Eq. (\ref{eq:cfh}), the correlations in Fig. \ref{fig:cfh} are much stronger for pions than for kaons or protons. Ultimately, this correlation leads to the balance function numerator. Because the definition of balance functions include a division by the yield of the specific hadron, the magnitude of balance functions for different species, e.g. $pp$ vs $\pi\pi$, are more similar than the correlations in Fig. \ref{fig:cfh}.

\begin{figure}
\centerline{\includegraphics[width=0.47\textwidth]{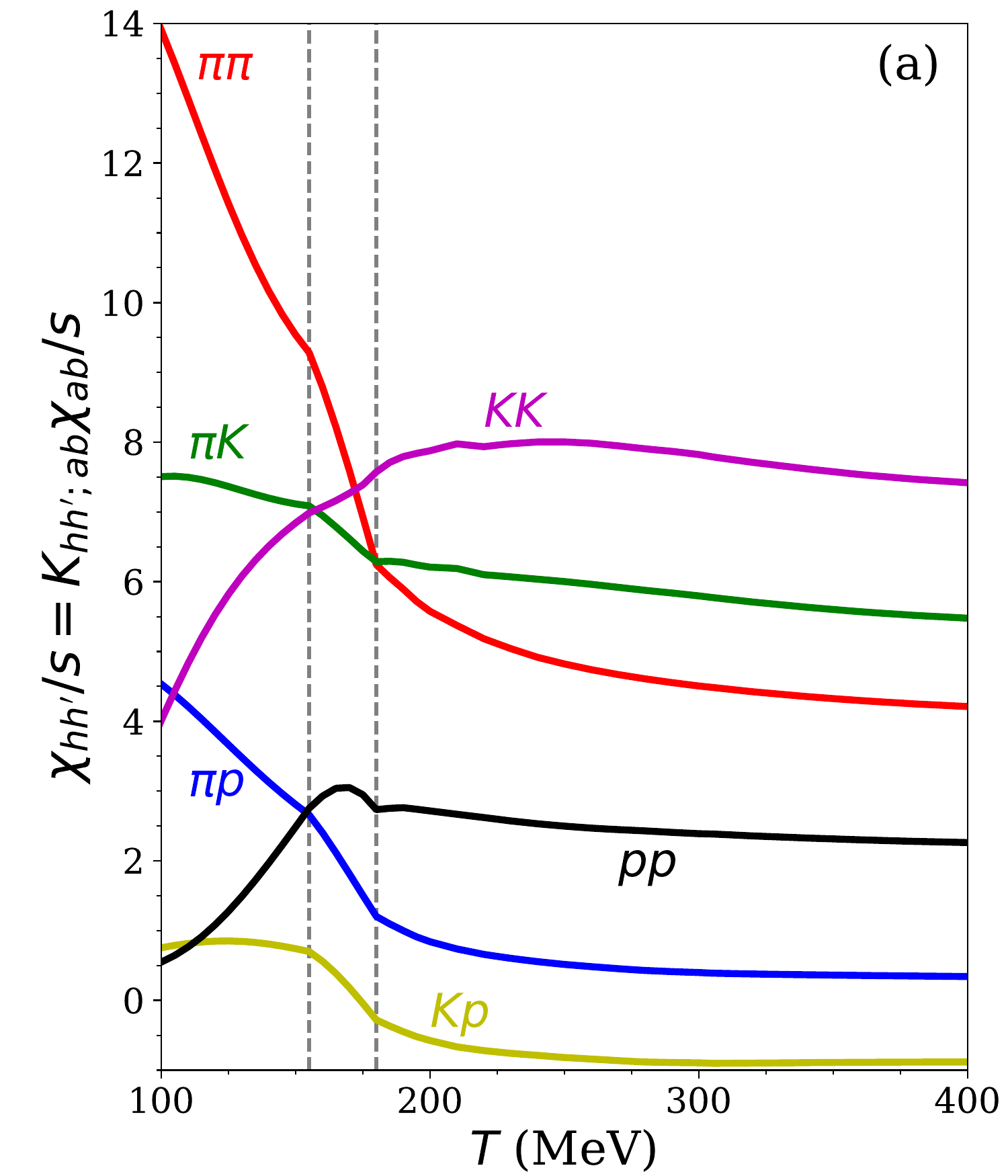}
\hspace{0.05\textwidth}
\includegraphics[width=0.4\textwidth]{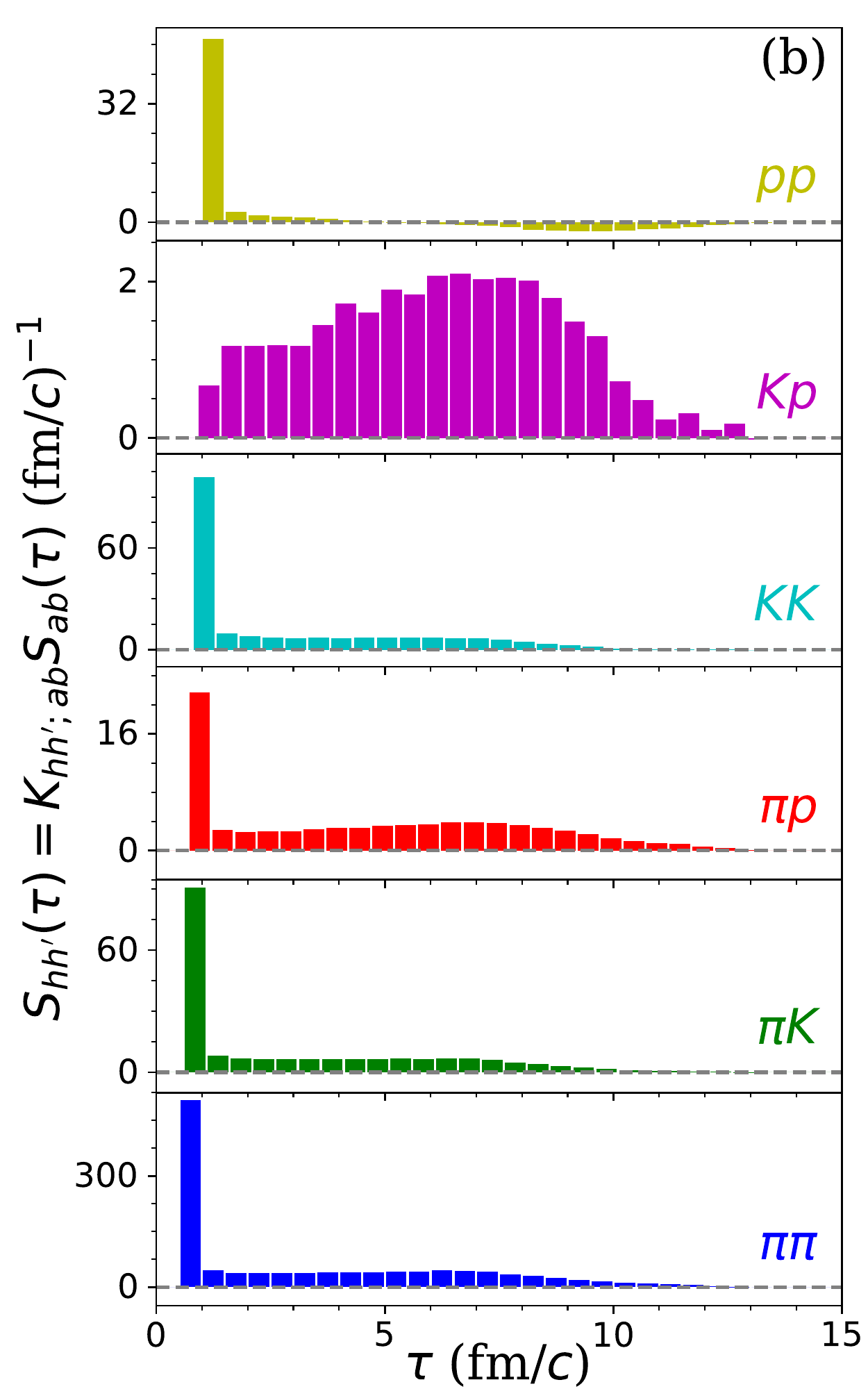}}
\caption{\label{fig:hsource}(color online)
The susceptibility, $\chi_{ab}(T)$, from Fig. \ref{fig:chilattice}, and the source function $S_{ab}(\tau)$, from Fig. \ref{fig:udssource}, are convoluted with the kernel $K_{hh';ab}$ to define $S_{hh'}$ and illustrate the temperatures and times at which correlations of specific hadron species are seeded.}
\end{figure}

The kernel defined in Eq. (\ref{eq:cfh}) can also be convoluted with the susceptibility or the source function,
\begin{eqnarray}
S_{hh'}(\tau)&\equiv&\sum_{ab}K_{hh';ab}S_{ab}(\tau),\\
\nonumber
\chi_{hh'}(\tau)&\equiv&\sum_{ab}K_{hh';ab}\chi_{ab}(\tau).
\end{eqnarray}
These quantities provide insight into the times and temperatures where the correlations that lead to final-state hadronic correlations for specific species, $hh'$, are seeded. They are illustrated in Fig. \ref{fig:hsource}. One can see that $S_{pp}$ has a large positive contribution at small $\tau$ and a negative contribution at large $\tau$ due to the reduction of baryon number in the hadronic phase due to the high thermodynamic penalty due to the high baryon masses. 

The correlation function $C'_{hh'}(x_1,x_2)$, graphed as a function of relative spatial rapidity in Fig. \ref{fig:cfh}, drives the final-state observed correlation. However, final-state correlations continue to develop during the hadronic stage. Balancing charges continue to diffuse away from one another in coordinate space, and resonances can decay or annihilate with one another, which represents an additional source to the correlation. For this study, treatment of the hadronic stage is confined to a Monte Carlo treatment of decays described in Sec. \ref{sec:method}. This affects the correlation in two ways. First, when a given hadron decays, there is a correlation between the products of that decay. For example, the decay of $\rho_0$ leads to a positive correlation between the $\pi^+\pi^-$ products. Second, if two balancing charges, $a$ and $b$, produce hadrons $h_a$ and $h_b$, via the Monte Carlo mechanism described in Sec. \ref{sec:method}, those two hadrons might well be unstable and decay \cite{Bozek:2003qi}. The final-state products are then evaluated for their contribution to the correlation function in lieu of their parent products. This can significantly affect the final-state correlations. For example, at the hyper-surface there may be a correlation between positive kaons and lambda hyperons because of the anti-strange quark in the $K^+$ and the strange quark in the lambda. The lambdas can undergo weak decay and produce protons, which then leads to a positive correlation between positive kaons and positive protons. This then leads to a negative contribution to $C_{Kp}$, and because it was originally seeded by the $ss$ correlations, which were seeded early, it is broad in $\Delta\eta$. To gain an idea of the role of feed-down corrections to the kernel, Fig. \ref{fig:chitrans} also presents the kernel including the effects of decays,
\begin{eqnarray}
\label{eq:feeddown}
\tilde{K}_{hh';ab}&=&\sum_{gg'}w_{gh}w_{g'h'}K_{gg';ab},
\end{eqnarray}
where $w_{gh}$ describes the average number of hadrons of type $h$ produced by the decay of a hadron of type $g$. Fig. \ref{fig:chitrans} shows how the influence of $ss$ correlations is spread over several other species by decays, primarily weak decays.


\section{Results}
\label{sec:results}

As described in Sec. \ref{sec:method}, once the sampling charges are converted into hadrons, the created hadron pairs are then decayed, with their products binned to create the numerator of the balance function, $N_{hh'}(\Delta y)$. The numerator also includes contributions from hadronic decays of an uncorrelated background. Balance functions, $B_{hh'}(\Delta y)$ were then created by dividing the numerator by the mean multiplicity of type $h$ hadrons. By tagging pairs of correlated particles, the uncorrelated background was strongly suppressed, which makes it possible to  accurately calculate $B_{hh'}(\Delta y)$ with only a few minutes of CPU time. By accounting for the experimental acceptance, as also described in Sec. \ref{sec:method}, the balance functions can be directly compared to data. Unlike much of the theoretical studies of fluctuations in heavy-ion collisions, which are based on heuristic arguments about choosing the right window in rapidity, to correspond to the charge fluctuation at a given stage of the collision, these calculations make strict quantitative predictions. These predictions cover a dozen bins of $\Delta y$, for each of six charge balance functions indexed by hadron species. For each balance function, details of the shape and magnitude should be reproduced if the model accurately depicts the chemical evolution, diffusion, and flow of the reaction. Here, we consider only balance functions binned by relative rapidity, but future studies might perform analyses based on bins of relative azimuthal angle, invariant relative momentum, or of the pair's average rapidity, average transverse momentum, or of the average azimuthal angle relative to the reaction plane. 

To date, only the STAR Collaboration at RHIC has presented charge balance functions indexed by identified hadronic species. Preliminary results exist for $\pi\pi$, $KK$, $pK$ and $pp$ binned by relative rapidity. There is no reason not to expect that $\pi K$ or $\pi p$ could be analyzed from previous experimental runs. If the goal is to determine $C_{ab}(\Delta\eta)$ from $B_{hh'}(\Delta y)$, and given there are only four independent correlations indexed by charges, $uu$, $ud$, $us$ and $ss$, only four sets of hadronic species, $hh'$, should be sufficient to determine $C_{ab}$. However, more measurements are always welcome, and would certainly enhance the confidence with which one could de-convolute $B_{hh'}$ to determine $C_{ab}$. In addition to studying all six hadronic combinations studied here, one could imagine identifying other species, such as lambdas, should experimental statistics become sufficient.

\begin{figure}
\centerline{\includegraphics[width=0.4\textwidth]{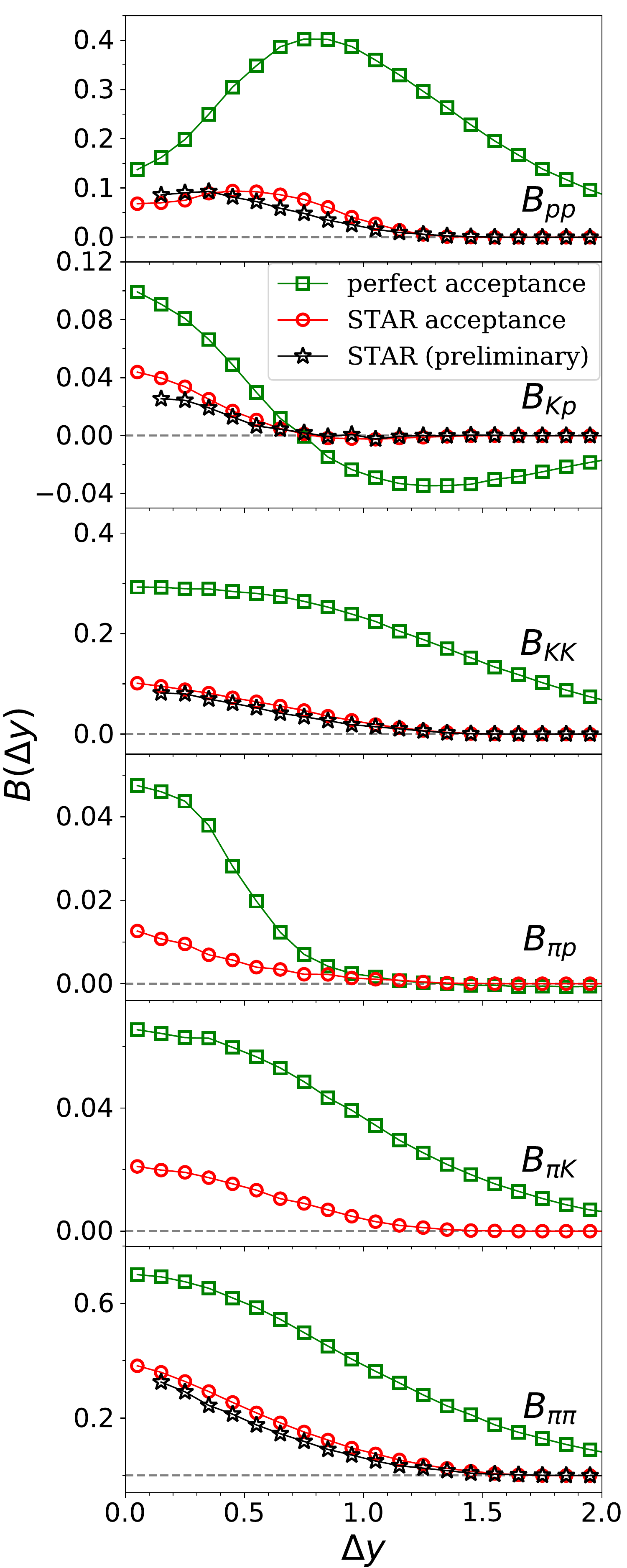}}
\caption{\label{fig:generalperfect}(color online)
Calculations for default parameters are shown with and without corrections for STAR acceptance. Applying the filter indeed brings the balance function into the neighborhood of preliminary measurements from STAR for central Au+Au collisions at $\sqrt{s}_{nn}=200$ GeV. Experimental results are reproduced at the $\sim$20\% level. The limited acceptance of STAR significantly restricts the insight one might possibly gain from a broader-acceptance device.}
\end{figure}

Figure \ref{fig:generalperfect} shows the balance functions calculated from the hydrodynamic model here, with and without the STAR acceptance filter. For each $\pi,K,p$ hadron measured in the STAR TPC (Time Projection Chamber) the chance that an associated particle, i.e. one related by charge conservation to a recorded particle, is also within the acceptance and is measured and identified is approximately 25\%. Thus, the charge balance functions are dampened by that same factor compared to calculations assuming a perfect acceptance. The balance function is completely damped as the width, $\Delta y$, approaches the maximum width of the detector. For such pairs, both particles must be emitted at the edge of the rapidity coverage, which forces the balance functions to zero. The maximum width is $\Delta y=1.8$ due to the pseudo-rapidity coverage of $\pm 0.9$. However, effectively, the damping is even stricter because particles with finite pseudo-rapidity $\eta=0.9$ maps to a lower rapidity. Thus, the acceptance more strongly dampens the proton and kaon balance functions than it does the $\pi\pi$ balance function because protons and kaons move more slowly than pions. 

Preliminary results from the STAR Collaboration \cite{Wang:2012jua} from central (0-5\% most central) Au+Au collisions at $\sqrt{s}_{nn}=200$ GeV. These are shown alongside the calculations in Fig. \ref{fig:generalperfect}. Model calculations were performed for default parameters. Due to baryon annihilation in the hadronic phase, the pion, kaon and baryon yields cannot be simultaneously matched by assuming chemical equilibrium at one temperature \cite{Pan:2014caa,Steinheimer:2012bn,Steinheimer:2017vju,Steinheimer:2012rd,Becattini:2012sq,Becattini:2012xb,Karpenko:2012yf}. First, the decoupling temperature was set to 140 MeV. This is a reasonable temperature for reproducing the baryon yield, but may be too low to reproduce kaon yields. The initial width of the charge correlations when the hydrodynamic treatment is begun was set at $\sigma_0=0.25$ and the diffusion constant used lattice values. The calculations roughly match the data, at the levels of 10-20\%. Significant discrepancies remain, which will be investigated below. 

\begin{figure}
\centerline{
\includegraphics[width=0.33\textwidth]{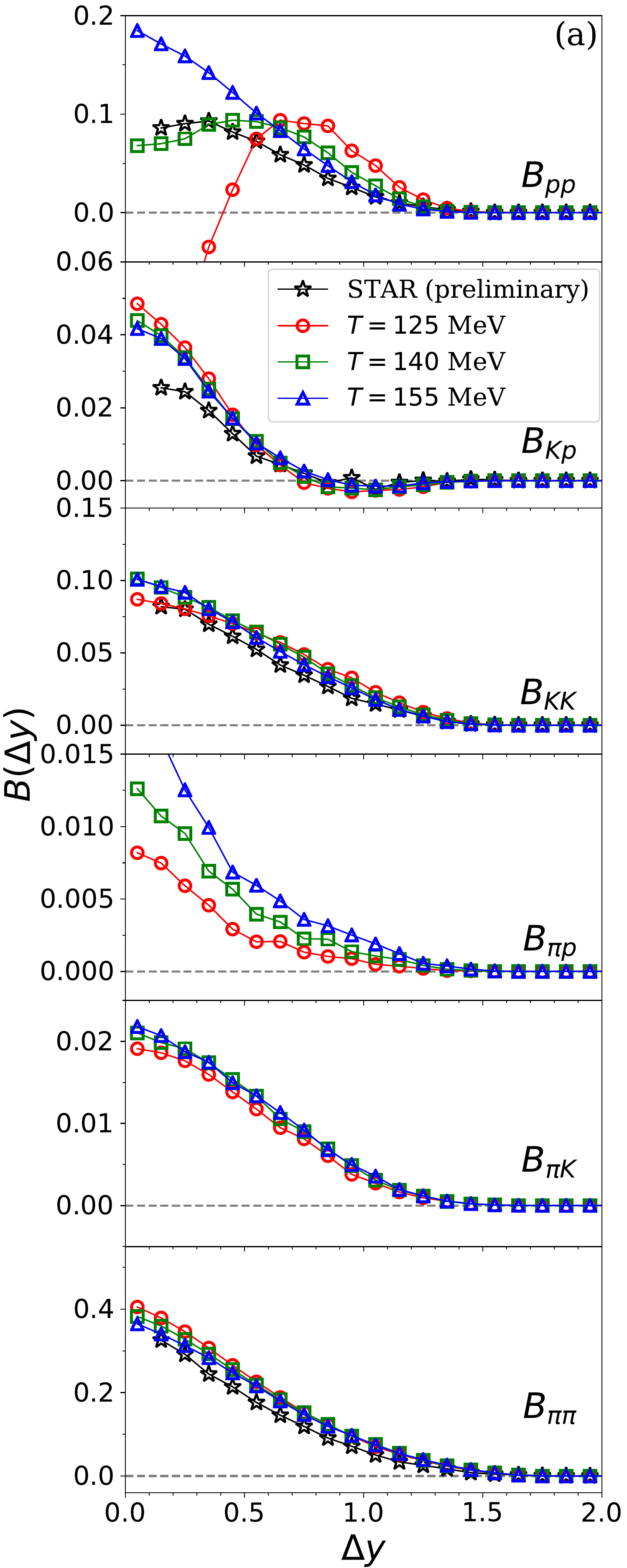}\hspace*{0.005\textwidth}
\includegraphics[width=0.33\textwidth]{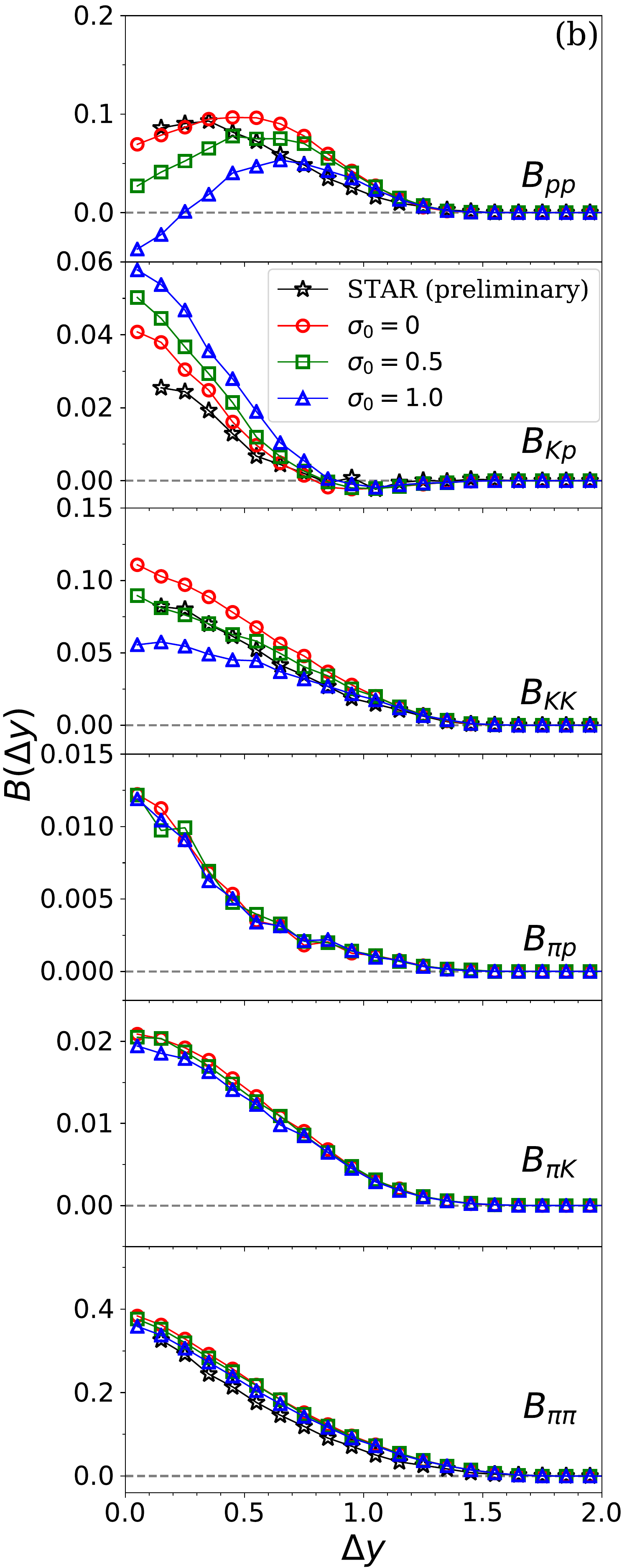}\hspace*{0.005\textwidth}
\includegraphics[width=0.33\textwidth]{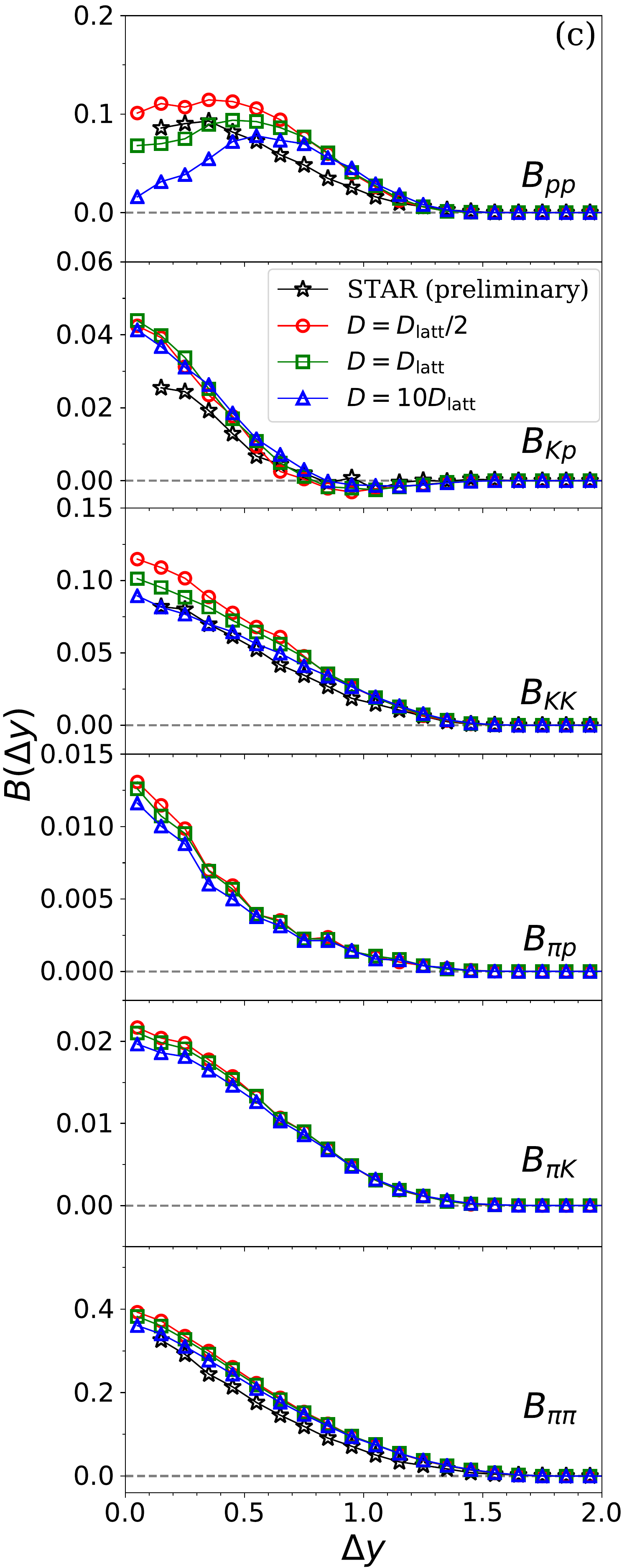}}
\caption{\label{fig:generalvarypars}(color online)
The left-side panels (a) show charge balance function calculations where the decoupling temperature is varied. For low temperatures, the fall in the baryon yield is accompanied by a negative contribution to the source function for $S_{aa}$ at late times. This then provides a dip in the balance function at small relative rapidity. Varying the width of the initial correlations, $\sigma_0$, broadens the correlations $C'_{aa}$ at early times. This mainly affects the balance functions involving protons and kaons. Finally, increasing the diffusion constant also more strongly broadens the correlations driven by early sources, such as $B_{KK}$ and $B_{pp}$.
}
\end{figure}

To gain insight into the sensitivity of the charge balance function to quantities that affect the production and diffusion of charge throughout the reaction region, three parameters were varied relative to the default calculation presented in Fig. \ref{fig:generalperfect}, and results are displayed in Fig. \ref{fig:generalvarypars}. First, the left-side panels demonstrate the sensitivity to variation of the decoupling temperature. Calculations for all six combinations of hadronic species were performed for three values of the decoupling temperature, $T_f=155, 140$ and 125 MeV. Of the six balance functions, those involving protons are the most affected because the yields of protons falls precipitously in calculations, like these, where chemical equilibrium is attained. The fall of the proton yield at low temperature requires that the components of the source function which are related to $C'_{pp}$ turn negative at large times, as shown in Fig. \ref{fig:hsource}. This leads to a dip in the $pp$ balance function at small relative rapidity. For lower decoupling temperatures, this dip becomes increasingly pronounced. For a perfect detector, the net area underneath the balance function curves should stay mainly independent of the decoupling temperature, or the yields, because for every observed proton, there should be either one more anti-baryon or one more baryon. To balance the stronger dip at $\Delta y\sim 0$ of the balance functions for lower $T_f$, the additional strength might well appear at larger relative rapidity, where observation is inaccessible due to the limited coverage of STAR.

The middle panels of Fig. \ref{fig:generalvarypars} demonstrates the sensitivity of the balance function to the width of the initial correlation, $C'(\Delta\eta,\tau=\tau_0)$. Because protons and kaons are relatively more sensitive to source functions at early times, they are also more sensitive to the initial width, $\sigma_0$. Increasing $\sigma_0$ from zero to one unit of relative rapidity moves strength of the $pp$ balance function away from the measured region. Sensitivities to changes of the diffusion constant are shown in the right-side panels of Fig. \ref{fig:generalvarypars}. Halving the diffusion constant more strongly constrains the balancing charge to smaller relative rapidity. Also shown are the effects of increasing the diffusion constant ten-fold. Increasing the diffusion is similar to what happens when $\sigma_0$ is increased because both effects mainly spread out the response to early-time elements of the source function, and thus both mainly affect the balance functions involving protons or kaons. In the future, one might analyze balance functions binned by relative azimuthal angle. In addition to being a sensitive test of radial flow \cite{Bozek:2004dt}, these balance functions should be sensitive to the diffusion constant, but not so sensitive to $\sigma_0$. 


\section{Summary and Outlook}
\label{sec:summary}

Charge fluctuations represent a well-posed observable that provides the best insight into a system's chemistry. Due to the evolving and finite-size nature of heavy-ion reaction regions, charge correlations represent the best chance of discerning insight into the local fluctuations, and into the evolving chemistry of a heavy-ion reaction. Here, for the first time, the evolution of the charge correlation function has been calculated based on a realistic hydrodynamic evolution and with realistic sources. The equation of state, that drives the hydrodynamics, the sources of the charge correlation, and the evolving diffusion constant are all taken from lattice calculations. A space-time description of the three-by-three charge correlation function is evaluated throughout the hydrodynamic phase, and the correlations are projected onto correlations with hadronic indices, i.e. charge balance functions. Charge balance functions were calculated for all combinations of charged pions, kaons and protons. These six balance functions were binned as a function of relative rapidity. Results are shown to be sensitive to the decoupling temperature, diffusion constant and to the initial width of the correlation at the beginning of the hydrodynamic evolution. 

The discrepancy between STAR measurements and the model was at the 20\% level for the default parameter set. Adjusting the three parameters mentioned above seemed able to change many of the balance functions at that level, but it appears that the data will not be reproduced at a high level of accuracy by adjusting these three parameters alone. The observed $pp$ balance function appears narrower than any of the calculations. It is difficult to see how one could reduce the strength for $\Delta y\sim 0.8$ from the model to fit the experimental $pp$ balance function without worsening the fits of other balance functions. The $\pi\pi$ balance function was also found to be modestly broader than the experimental one for all the parameter variation explored here. In fact, the $\pi\pi$ balance function was remarkably insensitive to any change in the three parameters described above.

It is possible that one could find a satisfactory reproduction of the experimental data by adjusting other parameters or by changing other model features. For example, one might increase the viscous distortion of the spectra. In the calculations, the viscous distortion was determined by the off-diagonal elements of the stress-energy tensor. However, those distortions were small because at the end of a central Au+Au collision the size of the transverse velocity gradients have grown close to the size of the longitudinal velocity gradient. Thus, even with a higher viscosity, the distortion might stay small. Another possibility would be to treat the initial width of the $ss$ and $uu$ or $dd$ correlations as separate parameters. 

However, fine-tuning parameters to reproduce data should not be done until the model is sufficiently sophisticated that, somewhere within its parameter space, it includes all the features whose inclusion or variation might change results at the level to which one wishes to reproduce the data. Even though these calculations are far more realistic than previous calculations, serious shortcomings remain. First, and most importantly, the hadronic stage is poorly treated. Currently, hadrons are immediately emitted from the hydrodynamic stage onto the final (aside from decays) state. Due to the loss of both chemical and local kinetic equilibrium, the standard way to treat the hadronic stage has been through microscopic simulations. The calculations shown here required little numerical expense (only a few minutes for a given point in parameter space) because one could tag those particles originally coming from the same source point. Once an emitted particle, $h$, collides with third bodies, the correlation spreads to a large number of particles, plus those final-state trajectories altered by the existence of $h$. Thus, evolving two-particle correlations through a hadron cascade is challenging. Given that the chemistry, e.g. baryon annihilation, evolves significantly during this stage, and in ways not particularly well reproduced by equilibrium models, this remains a priority going forward. 

By adding a microscopic simulation, one would expect a narrowing of the balance function, which should improve agreement with experiment. For example, if charges are created early and if the system broke up immediately, before collective flow developed, balance functions in azimuthal angle would be broad, extending all the way to $\Delta\phi=\pi$. Collisions, and the growth of strong collective flow focuses conserved charge pairs into a tight window. By adding more collisions, and more collective flow, by adding a hadronic simulation, one would expect further narrowing of the balance function. One would also expect further narrowing in relative rapidity when adding a hadronic simulation. This might be especially important for the $pK$ balance function. The narrow component is related to the source $us$ and $ds$ source functions, which develop late in the collision and have narrow widths in coordinate space. Thus, this peak is very sensitive to the thermal broadening, which might be significantly reduced by the additional cooling that might ensue in a hadronic cascade. However, it is difficult to estimate how much additional narrowing would ensue. As the system further evolves, collective flow increases, but so does the diffusive separation in coordinate space.

It should also be restated that the calculations here were based on the assumption of boost invariance, of equal numbers of particles and anti-particles, and of zero isospin. It is not clear how improving any of these would change the resulting balance functions, or if the changes would even be substantial. Even changes of a few percent might be important if they are combined with other changes of a few percent. Finally, it should be emphasized that the data is preliminary. Instead of dividing the numerator of the balance function by the yields to find a balance function, one could divide by the correlation by similar binnings of mixed events. One would then see that the strength of that correlation would be at the level of a few tenths of a percent. Thus, any unaccounted detector effects might significantly alter the balance function. As seen in Fig. \ref{fig:generalperfect} the final experimentally constructed balance functions are strongly damped by the experimental acceptance and efficiency. Any changes in the efficiency might also make a significant difference, so the authors look forward to seeing fully vetted results. 

Despite the challenges described above, the possibilities for this class of observable are exciting. Once one separates charge balance function by hadron species, it would appear that it becomes possible to image the final-state correlation $C'_{ab}$, expressed in spatial rapidity, using experimental balance functions binned by relative rapidity in momentum space, $B_{hh'}(\Delta y)$. Six  combinations $hh'$ are experimentally observable, which should enable imaging of the entire three-by-three correlation $C'_{ab}(\Delta \eta)$. Ultimately, one could analyze balance functions in terms of all six dimensions of relative and average momenta, similar to what happened with femtoscopic correlations at small relative momentum. 

If the imaging above is indeed accomplished, it means much more than the reproduction of an experimental measurement, even a rich one. These analyses provide potential insight and answers to some of the field's most basic questions: Did we produce a chemically equilibrated quark-gluon plasma? If so, how and when was the charge produced? And, can we extract an otherwise unobtainable bulk property of the quark-gluon plasma -- the diffusion constant for light quarks?


\appendix*
\section{Generating Hyper-Surface Elements from Hydrodynamic History}

The Cooper-Frye formula \cite{Cooper:1974mv} is used to generate a list of emitted particles from a given hyper-surface element, $d\Omega_\mu$. When these elements are time-like, one can boost to a frame where the element has only a time component and $d\Omega_0$ is a differential volume in that frame. If the element is space-like, one can boost and rotate to a frame where $d\Omega_z$ is the only component, and represents a differential area with its normal pointed in the $z$ direction, multiplied by a differential time.

To generate hadrons, either to represent the emission or to feed a hadronic simulation, one starts with a list of hyper-surface elements. For each element, the information needed is: a) the four components of $d\Omega_\mu$ and b) the collective velocity $u^\mu$ at that point. Typically, one also records the position at the center of the element if one plans to use the particles to feed a simulation. Also, if the hyper-surface is not defined by a fixed temperature and chemical potential, one would also record those values. Hydrodynamic histories are usually stored in rectangular meshes. For our case, we assume boost invariance, so the information is three-dimensional, divided into bins of proper time $\tau$, and the two transverse coordinates $x$ and $y$. The step sizes are $d\tau$ and $dx=dy$. It is straightforward to extend the ideas below to meshes with varying step sizes or to four-dimensional meshes.

Formally, a hypersurface is defined by some function $C(T(r),\mu(r))=0$, where $r$ is a four-vector describing the space-time position, $T$ is a temperature and $\mu$ is the chemical potential. For $C<0$, one is inside the hydrodynamic region and for $C>0$, one is in the hadronic realm. The most common choice for systems with equal numbers of particles and antiparticles is $C=T_f-T$. The hyper-surface is defined as
\begin{eqnarray}
\label{eq:hyperfindC}
d\Omega_\mu&=&d^4r ~\partial_\mu~\Theta(C(r)),
\end{eqnarray}
where $\Theta$ is a step function. For $C(r)=T_f-T(r)$,
\begin{eqnarray}
d\Omega_\mu&=&-d^4r~\delta(T(r)-T_f)\partial_\mu T(r).
\end{eqnarray}
One can use the delta function to reduce the differential by one dimension,
\begin{eqnarray}
\label{eq:3criteria}
d\Omega_\mu&=&-\left.\frac{\tau\Delta\eta~ dx~dy}{\partial_\tau T}\right|_{\tau_f}\partial_\mu T(r),\\
\nonumber
&=&-\left.\frac{\tau\Delta\eta~ dy~d\tau}{\partial_x T}\right|_{x_f}\partial_\mu T(r),\\
\nonumber
&=&-\left.\frac{\tau\Delta\eta~ dx~d\tau}{\partial_y T}\right|_{y_f}\partial_\mu T(r).
\end{eqnarray}
Any of the three expressions above (would be four if there weren't invariance in $\Delta\eta$) suffice in principle, but to increase accuracy one may choose one of the three based on the particular situation.

The temperature is stored at discrete points, $\tau_i,x_j,y_k$, and a differential element is defined by the eight points at the edges of the cubic differential volume, $r_{i,j,k}=\tau_i,x_j,y_k,r_{i+1,j,k}$, $r_{i,j+1,k}$, $r_{i,j,k+1}$, $r_{i,j+1,k+1},r_{i+1,j,k+1}$, $r_{i+1,j+1,k}$, $r_{i+1,j+1,k+1}$. For the first expression in Eq. (\ref{eq:3criteria}), one checks to see whether both $T$ evaluated at the later time, $i+1$, is below $T_f$ and $T$ evaluated at the earlier time, $i$, is above $T_f$. These two temperatures are
\begin{eqnarray}
T_{\tau-}&=&\frac{1}{4}\left(T_{i,j,k}+T_{i,j+1,k}+T_{i,j,k+1}+T_{i,j+1,k+1}\right),\\
T_{\tau+}&=&\frac{1}{4}\left(T_{i+1,j,k}+T_{i+1,j+1,k}+T_{i+1,j,k+1}+T_{i+1,j+1,k+1}\right).
\end{eqnarray}
If a differential volume satisfies this criteria, i.e. $(T_f-T_{\tau+})(T_f-T_{\tau-})<0$, the volume is added to the list of hyper-elements. The derivatives, $\partial_\mu T$, are found by averaging the four values at the points just above/below the center point of the sphere, evaluated in the center of the faces of each sphere. For this case,
\begin{eqnarray}
\partial_\tau T&=&\frac{T_{\tau+}-T_{\tau-}}{d\tau},\\
\nonumber
\partial_x T&=&\frac{T_{x+}-T_{x-}}{dx},\\
\nonumber
\partial_y T&=&\frac{T_{y+}-T_{y-}}{dy}.
\end{eqnarray}
Here, $T_{x-}$, $T_{x+}$, $T_{y-}$ and $T_{y+}$ are defined in similar fashions to $T_{\tau-}$ and $T_{\tau+}$. Using the first criteria in Eq. (\ref{eq:3criteria}),
\begin{eqnarray}
d\Omega_0&=&\tau\Delta\eta dxdy,\\
\nonumber
d\Omega_x&=&\tau\Delta\eta d\tau dy \frac{\partial_x T}{\partial_\tau T},\\
\nonumber
d\Omega_y&=&\tau\Delta\eta d\tau dx \frac{\partial_y T}{\partial_\tau T}.
\end{eqnarray}
For these last two components the ratios of derivatives represent the inverse speeds at which the surfaces collapse, and $|d\vec{\Omega}|= 
d\Omega_0/v_\Omega$, where $v_\Omega$ is the speed at which the surface moves inward, anti-parallel to the temperature gradient. The direction of $d\vec{\Omega}$ is also anti-parallel to the temperature gradient, typically outward toward the hadronic phase. If the surface collapses instantaneously, that speed is infinite and $d\Omega_0$ is the only non-zero component of the hyper-volume.

Rather than choosing the first expression in Eq. (\ref{eq:3criteria}), one could use the second, or third. For a static surface, the derivative $\partial_\tau T=0$, and $d\Omega_x$ and $d\Omega_y$ are undefined. To avoid this, one uses the local temperature gradients for each volume element to choose which of the three criteria to apply. Comparing, $\delta T_\tau=d\tau|\partial_\tau T|$, $\delta T_x=dx|\partial_xT|$, $\delta T_y=dy|\partial_yT|$, one picks the largest of the three, then correspondingly chooses from three criteria.  With this choice, the hyper-surface will be represented by the largest number of hyper-elements, e.g. more elements with smaller volumes, and thus should be most accurate. In the code used here, the criteria was slightly altered. If $\delta T_\tau>dx|\nabla T|$, one would use the first criteria. Otherwise, one would base the choice on $\delta T_x$ vs $\delta T_y$. For most of the emission the latter two criteria were typically used, and for the last few fm/$c$, the first criteria came into play. Because the hydrodynamic evolution was written with much smaller time steps than spatial steps, $d\tau=0.02$ fm/$c$ and $dx=dy=0.1$ fm/$c$, the first criteria did not come into play until the hyper surface was collapsing much faster than the speed of light. Adjusting the criteria for choosing between the three expressions in Eq. (\ref{eq:3criteria}) had only a negligible effect.

To test the algorithm described above, an artificial space-time evolution was considered. For this case, the shape of the hyper-surface was taken to be circular, by defining a temperature and collective velocity by
\begin{eqnarray}
\label{eq:fakehydro}
R(\tau)&=&R_0-v_s(\tau-\tau_0),\\
\nonumber
T(r,\tau)&=&T_f-T_s(r-R(\tau))/R_0,\\
\nonumber
u_x&=&u_R\frac{r}{R(\tau)}\cos\phi,\\
\nonumber
u_y&=&u_R\frac{r}{R(\tau)}\sin\phi,\\
\nonumber
u_z&=&\sinh\eta\sqrt{1+u_x^2+u_y^2},\\
\nonumber
u_0&=&\cosh\eta\sqrt{1+u_x^2+u_y^2}
\end{eqnarray}
with $R_0=10$ fm, $\tau_0=0.6$ fm/$c$, $T_s=100$ MeV, $T_f=155$ MeV and $u_R=1$. This hyper-surface collapsed at the speed of light, $v_s=1$. Spectra were calculated both by a semi-analytic treatment (still involving a integral over time) to one where the temperature and collective velocities were written to a mesh with resolution $d\tau$ and $dx=dy$. 

For the semi-analytic treatment, the hyper surface elements are found using Eq. (\ref{eq:hyperfindC}),
\begin{eqnarray}
d\Omega_\mu&=&\tau d\eta d\phi~ r~ d\tau~ \left.\frac{\partial_\mu T}{\partial_r T}\right|_{r=R(\tau)},\\
\nonumber
d\Omega_0&=&\cosh\eta\tau d\eta d\phi d\tau v_s R(\tau),\\
\nonumber
d\Omega_x&=&\tau d\eta d\phi d\tau R(\tau) \cos\phi,\\
\nonumber
d\Omega_y&=&\tau d\eta d\phi d\tau R(\tau) \sin\phi,\\
\nonumber
d\Omega_z&=&\sinh\eta\tau d\eta d\phi d\tau v_s R(\tau).
\end{eqnarray}

The spectra for particles of mass $m$ are then found by integrating the Cooper-Frye Formula \cite{Cooper:1974mv},
\begin{eqnarray}
\label{eq:cooperfrye}
E_p\frac{dN}{d^3p}&=&
\int d\Omega_\mu p^\mu e^{-u\cdot p/T_f},\\
\nonumber
&=&\int d\eta \tau d\tau d\phi R(\tau) (v_s E_p\cosh\eta +p\cos\phi)\exp\{-\beta E\sqrt{1+u_R^2}\cosh\eta+\beta pu_R \cos\phi\}.
\end{eqnarray}
The integrals over $\eta$ and $\phi$ can be performed analytically,
\begin{eqnarray}
\label{eq:cfanal}
E_p\frac{dN}{d^3p}&=&\frac{(2s+1)}{2\pi^2}\int_{\tau_0}^{\tau_0+R_0/v_s} d\tau~\tau R(\tau)\\
\nonumber
&&\cdot\left\{v_s EK_1(\beta E\sqrt{1+u_R^2})I_0(\beta p u_R)+p_\perp K_0(\beta E \sqrt{1+u_R^2})I_1(\beta p u_R)\right\}.
\end{eqnarray}
Here, the intrinsic spin is $s$ and $\beta=1/T$.

\begin{figure}
\centerline{\includegraphics[width=0.4\textwidth]{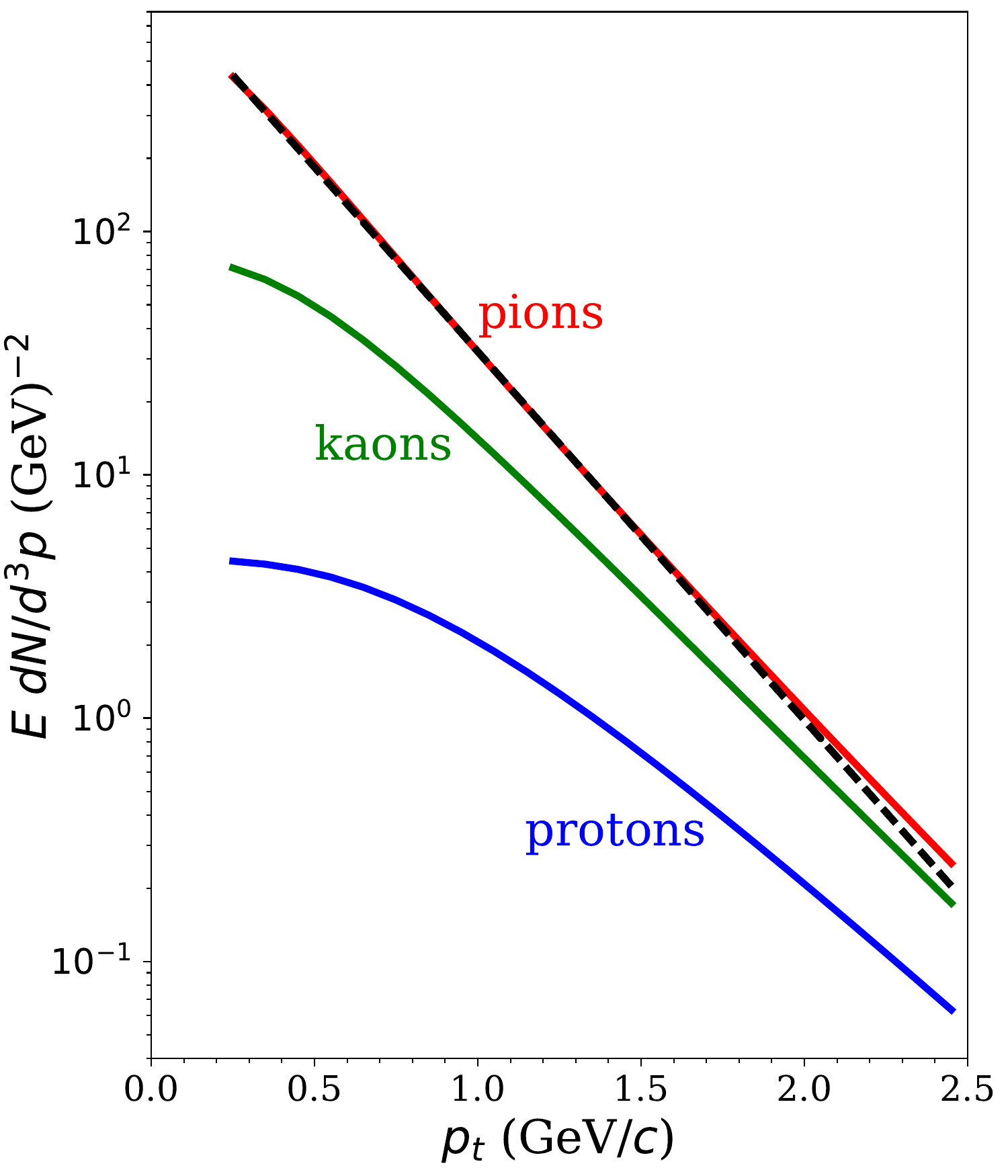}}
\caption{\label{fig:cftest}(color online)
Spectra for the simple parameterization of the hydrodynamic history, described Eq. (\ref{eq:fakehydro}), are calculated from a nearly analytic formula, Eq. (\ref{eq:cfanal}), and from reading the history from a mesh and applying the algorithm described here. The nearly-analytic (lines) and the spectra calculated from a discretized mesh (points) agree well.}
\end{figure}

Spectra for pions, kaons and protons were calculated by numerically integrating the expression in Eq. (\ref{eq:cfanal}). They were then compared to results obtained by reading the hydrodynamic history, described in Eq. (\ref{eq:fakehydro}), from a mesh, then following the algorithm to generate hyper-surface elements described above. The size of the mesh, $d\tau=0.02$ fm/$c$, $dx=dy=0.1$ fm/$c$, was the same as what was used to record the real hydrodynamic evolutions. Figure \ref{fig:cftest} shows that the spectra indeed match well, with discrepancies below one tenth of one percent. The test was repeated with a coarser mesh, and the agreement was still at the level of one or two tenths of a percent.

In addition to testing the spectra, the quantity
\begin{eqnarray}
v_4\equiv\langle \cos 4\phi\rangle,
\end{eqnarray}
was calculated using the hydrodynamic mesh. Here, the $\langle ...\rangle$ denotes an average over emitted particles. Because the surface ostensibly has cylindrical symmetry, $v_4$ should be zero. However, the rectangular mesh explicitly breaks the symmetry, so the measure of $v_4$ provides a test of the accuracy of the procedure. For the same resolution used here, $|v_4|$ was on the order of $10^{-5}$. 

This method applied here is fast and easy to implement, but it is by no means the most sophisticated. In \cite{Huovinen:2012is} the contribution from each cell is evaluated in detail, and the surface through the interior of a cell is estimated, which may be larger or smaller depending on the degree to which the hyper-surface clips the cell. Because the method presented here seems to work at better than the 0.1\% level, it is sufficient for this example, but the methods of \cite{Huovinen:2012is} might be warranted for systems with coarse grids, or with small scale temperature variations.

\begin{acknowledgments}
This work was supported by the Department of Energy Office of Science through grant number DE-FG02-03ER41259 and through grant number DE-FG02-87ER40328. The authors express their gratitude to Claudia Ratti for providing the data tables for the lattice calculations of the charge susceptibility.
\end{acknowledgments}


\end{document}